\documentclass[10pt]{article}

\usepackage[english]{babel}
\usepackage{amsthm}
\usepackage{amssymb}
\usepackage{mathtools}

\newtheorem{theorem}{Theorem}[section]
\newtheorem{lemma}[theorem]{Lemma}
\newtheorem{proposition}[theorem]{Proposition}
\newtheorem{corollary}[theorem]{Corollary}
\newtheorem{remark}[theorem]{Remark}
\newtheorem{definition}[theorem]{Definition}
\makeatletter

\@addtoreset{equation}{section}
\usepackage{color}
\usepackage{graphicx}

\makeatother
\usepackage{newlfont}
\def \n {\noindent}
\begin{document}
\begin{center}
{\color{blue}{\bf {\Large New Spectral Properties of Imaginary part }}}\\
{\color{blue}{\bf {\Large  of Gribov-Intissar Operator}}}
\end{center}
\begin{center}
{\bf Abdelkader Intissar$^{(1)}$}\\
\n {\it (1) Le Prador,129, rue du commandant Rolland, 13008 Marseille-France}.\\
\n E.mail: {\color{blue}abdelkader.intissar@orange.fr}\\
\end{center}
\begin{center}
\n {\bf{\large {\color{red} Abstract}}}
\end{center}

\n In 1998, we have given  in ({\color{blue}[14]} Intissar, A., Analyse de Scattering d'un op\'erateur cubique de Heun dans l'espace de Bargmann, Comm.Math.Phys.199 (1998) 243-256) the boundary conditions at infinity for a description of all maximal dissipative extensions in Bargmann space of the minimal Heun's operator.\\

 $\displaystyle{ H_{I} = z( \frac{d}{dz} + z)\frac{d}{dz}}$; $z \in \mathbb{C}$. The characteristic functions of the dissipative extensions have computed and some completeness theorems have obtained for the system of generalized eigenvectors. In ({\color{blue}[18]} Intissar, A, Le Bellac, M. and Zerner, M., Properties of the Hamiltonian of Reggeon field theory, Phys. Lett. B 113 (1982) 487-489) the non self-adjoint operator $ \lambda H_{I}$ where $\lambda \in \mathbb{R}$ is imaginary  part of the Hamiltonian of Reggeon field theory :\\

$$\displaystyle{H_{\mu, \lambda} = \mu z\frac{d}{dz} + i \lambda z( \frac{d}{dz} + z)\frac{d}{dz} \,\, \text{where} \,\, (\mu, \lambda) \in \mathbb{R}^{2} \,\, \text{and} \,\, i^{2} = -1}$$
\n The main purpose of the present work is to present some new spectral properties of right inverse $K_{0, \lambda}$ of  $H_{\lambda} = i\lambda H_{I}$ ($H_{\lambda}K_{0, \lambda} = I$) on negative imaginary axis and to study  the deficiency numbers of the generalized Heun's operator $\displaystyle{ H^{p,m} = z^{p}( \frac{d^{m}}{dz^{m}} + z^{m})\frac{d^{p}}{dz^{p}}}$ $ p, m = 1, 2, ....$. In particular, here we find some conditions on the parameters $p$ and $m$ for that $H_{I}^{p,m}$ to be completely indeterminate.It follows from these conditions that $H^{p,m}$ is entire of the type minimal.\\

\n {\bf Keywords:} Spectral theory; Gribov-Intissar operators; Non-self-adjoint operators; Cubic Heun's operator; Bargmann space ; Reggeon field theory.\\
\newpage

\n {\bf{\color{black}\Large{ 0 }}} {\bf{\color{red}\Large{Introduction}}}\\

\n We recall that the Gribov-Intissar Operator is defined by :\\

\begin{equation}
\displaystyle{H_{\mu,\lambda} = \mu A^{*}A + i\lambda A^{*}(A + A^{*} )A}
\end{equation}
 \n where $A$, $A^{*}$ are the annihilation and creation operators, $\mu$, $\lambda$ are real parameters and $ {\color{red}i^{2} = -1}$.\\
 
\n  This operator is considered to act on Bargmann space {\color{blue}[3]}:\\
\begin{equation}
 \displaystyle{ \mathbb{B} = \{\varphi : \mathbb{C}  \longrightarrow \mathbb{C} \quad \text{entire};\,\, \int_{\mathbb{C}}\vert \varphi(z) \vert^{2} e^{-\vert z \vert^{2}}dxdy < \infty\}}
 \end{equation}
 \n where its usual basis is given by\\
 \begin{equation}
  \displaystyle{e_{n} (z) = \frac{z^{n}}{\sqrt{n!}}; n \in \mathbb{N}}
  \end{equation}
 \n and  the annihilation operator $A$ and  the creation operator $A^{*}$ are defined  by\\
 \begin{equation}
 \displaystyle{A\varphi:= \frac{d\varphi}{dz} \quad \varphi \in \mathbb{B}}
 \end{equation}
  and \\
  \begin{equation}
  \displaystyle{A^{*}\varphi = z\varphi \quad \varphi \in \mathbb{B}}
  \end{equation}
  \n Action of $A$ and of $A^{*}$ on usual basis $(e_{n}(z))_{n \in \mathbb{N}}$ is given respectively by \\
 \begin{equation}
 \displaystyle{Ae_{n} = \sqrt{n} e_{n-1}, \,\, n \geq 1\quad \text{and} \,\, Ae_{0} = 0}
  \end{equation}
  
 \begin{equation} 
  \displaystyle{A^{*}e_{n} = \sqrt{n+1} e_{n+1}, \,\, n \geq 0}
 \end{equation}
 \n and the expressions of $H_{\mu, \lambda}$ and of $H_{\lambda}$ are  given respectively  by\\
 \begin{equation}
 \displaystyle{H_{\mu,\lambda}\varphi = \mu z\frac{d\varphi}{dz} + i\lambda(z\frac{d^{2}\varphi}{dz^{2}} + z^{2}\frac{d\varphi}{dz}) \quad \varphi \in \mathbb{B}}
 \end{equation}

 \begin{equation}
 \displaystyle{H_{\lambda}\varphi = i\lambda(z\frac{d^{2}\varphi}{dz^{2}} + z^{2}\frac{d\varphi}{dz}) \quad \varphi \in \mathbb{B}}
 \end{equation}

\n {\color{red}$\bullet_{4}$ } In {\color{blue}[10]} and {\color{blue}[11]} we have given a complete spectral analysis of  the following {\color{red}$\mathbb{C}$-symmetric} matrices which play an important role in Reggeon field theory\\
\begin{equation}
\n \mathbb{H}_{n}^{\mu,\lambda}= \left(
  \begin{array}{ c c c c c c c c c } 
     \mu & {\color{red}i}\lambda\sqrt{2}& 0 & .&.&.&.& \\ 
     {\color{red} i}\lambda\sqrt{2}  & 2\mu & {\color{red}i}\lambda 2\sqrt{3} &\ddots & .&.&.\\
    0 & {\color{red}i}\lambda2\sqrt{3} & 3\mu & {\color{red}i}\lambda3\sqrt{4}& \ddots &.&.\\
    . & \ddots & \ddots & \ddots & \ddots & \ddots & .&\\
     . & . & \ddots & \ddots & \ddots  & \ddots& 0 & \\
     . & . & . & \ddots & \ddots  & \ddots & {\color{red}i}\lambda (n-1)\sqrt{n} &\\
     .   &. & . &.& 0 &  {\color{red}i}\lambda (n-1)\sqrt{n} & n\mu& \\
  \end{array} \right).
  \end{equation}
  \n where $\mu$, $\lambda$ are real parameters and $ {\color{red}i^{2} = -1}$.\\
  \quad\\
 \n These {\color{red}$\mathbb{C}$-symmetric} matrices  approximate our unbounded operator $H_{\mu, \lambda}$. \\

\n {\color{red}$\bullet_{5}$} It was well known for several years that the eigenvalues of this operator are real and that recently we have shown the completeness of its generalized eigenvectors. For this topic see {\color{blue}[12]}\\

\n  Let \\
\begin{equation}
\displaystyle{\mathbb{B}_{0} = \{\varphi \in \mathbb{B} ; \varphi(0) = 0\}}
\end{equation}
\n Then it was well known that on $\displaystyle{\mathbb{B}_{0}}$, for $\psi \in \mathbb{B}_{0}$, an explicit inverse of $H_{\mu, \lambda}$ restricted on imaginary axis ; $y \in [0,  +\infty[$ is given (see proposition  9 {\color{blue} [13]} by\\
\begin{equation}
\displaystyle{K_{\mu,\lambda}:= H_{\mu, \lambda}^{-1} \psi(-iy) = \int_{0}^{\infty}\mathcal{N}_{\mu ,\lambda}(y, s)\psi(-is)ds}
\end{equation}
\n where \\
\begin{equation}
\displaystyle{\mathcal{N}_{\mu,\lambda}(y, s) = \frac{1}{\lambda s}e^{-\frac{s^{2}}{2} - \frac{\mu}{\lambda}s}\int_{0}^{min(y, s)}e^{\frac{u^{2}}{2} +\frac{\mu}{\lambda}u}du}
\end{equation}

\n or by taking $\lambda = 1$ (see  lemme 4 in {\color{blue} [1]}) 

\begin{equation}
\displaystyle{K_{\mu}:= H_{\mu, 1}^{-1} \psi(-iy) = \int_{0}^{\infty}\mathcal{N}_{\mu}(y, s)\psi(-is)ds}
\end{equation}
\n where \\
\begin{equation}
\displaystyle{\mathcal{N}_{\mu}(y, s) = \frac{1}{s}e^{-\frac{s^{2}}{2} - \mu s}\int_{0}^{min(y, s)}e^{\frac{u^{2}}{2} + \mu u}du}
\end{equation}

\n The kernel of operator $\displaystyle{H_{\mu, \lambda}^{-1}}$  is analytic with respect to $\mu$. The integral operator defined by this kernel extends into a compact operator on a space $L_{2}$ with weight, including for negative values of $\mu$.\\

\n In particular, we have\\
\begin{proposition} (Ando-Zerner) \\

\n Let $\displaystyle{L_{2}([0, \infty[, e^{- x^{2} - 2\frac{\mu}{\lambda}x}dx)}$ with $\lambda \neq 0$ be a space of square integrable functions with respect the measure $\displaystyle{e^{- x^{2} - 2\frac{\mu}{\lambda}x}dx)}$ then we have\\

\n (i) For all $\mu > 0$,  $\displaystyle{H_{\mu, \lambda}^{-1}}$ can be extended to a Hilbert-Shmidt operator of \\$\displaystyle{L_{2}([0, \infty[, e^{- x^{2} - 2\frac{\mu}{\lambda}x}dx)}$ to itself.\\

\n (ii) The map $\displaystyle{ \mu \longrightarrow K_{\mu, \lambda}}$ is analytic on $[0, +\infty[$ in Hilbert-Schmidt norm operators on $\displaystyle{L_{2}([0, \infty[, e^{- x^{2} - 2\frac{\mu}{\lambda}x}dx)}$.\\

\n (iii) For $\mu > 0$,  the smallest eigenvalue $\sigma_{0}(\mu)$ of $H_{\mu, \lambda}$ which simple can be extended to  a real positive analytical function and  creasing  with respect $\mu$ on entire real axis.\\
\end{proposition} 

\n {\color{red}{\bf Proof}}\\

\n See  proposition 9 {\color{blue}[13]}, lemma 4 and lemma 8 of Ando-Zerner in {\color{blue}[1]}.\\

\n $H_{\mu,\lambda}$ is not self-adjoint operator, nevertheless it has several properties analogous to those of the self-adjoint operators :\\

\n {\color{red} ($\bullet_{1}$)} In 1987, we have given  in {\color{blue}[13]} many spectral properties of $H_{\mu, \lambda}$ for $\mu > 0$ in particular the minimal domain of $\displaystyle{H_{\mu, \lambda}}$ coincides with its maximal domain, the positiveness of its eigenvalues , the existence of the smallest eigenvalue $\sigma_{0} \neq 0$ and an asymptotic expansion of its semigroup $\displaystyle{e^{-tH_{\mu, \lambda}} }$ as $t \longrightarrow + \infty$.\\
\n The minimal domain of $\displaystyle{H_{\mu, \lambda}}$ is given by:  \\
\begin{equation}
\displaystyle{ D(H_{\mu, \lambda}^{min}) = \{ \varphi \in \mathcal{B} , \exists \, p_{n} \in \mathcal{P} , p_{n} \longrightarrow \varphi , \exists \, \psi \in \mathcal{B} ; H_{\mu, \lambda}p_{n} \longrightarrow \psi \}}
\end{equation}

\n For $\mu = 0$, the matrices \\

\begin{equation}
\n \mathbb{H}_{n}^{0,\lambda}= \left(
  \begin{array}{ c c c c c c c c c } 
     0 & {\color{red}i}\lambda\sqrt{2}& 0 & .&.&.&.& \\ 
     {\color{red} i}\lambda\sqrt{2}  & 0 & {\color{red}i}\lambda 2\sqrt{3} &\ddots & .&.&.\\
    0 & {\color{red}i}\lambda2\sqrt{3} & 0 & {\color{red}i}\lambda3\sqrt{4}& \ddots &.&.\\
    . & \ddots & \ddots & \ddots & \ddots & \ddots & .&\\
     . & . & \ddots & \ddots & \ddots  & \ddots& 0 & \\
     . & . & . & \ddots & \ddots  & \ddots & {\color{red}i}\lambda (n-1)\sqrt{n} &\\
     .   &. & . &.& 0 &  {\color{red}i}\lambda (n-1)\sqrt{n} & 0& \\
  \end{array} \right).
  \end{equation}
  
  \n approximate the limit case operator: \\
  \begin{equation}
  \displaystyle{H_{\lambda} = H_{0, \lambda} =  i\lambda A^{*}(A + A^{*} )A = i\lambda [z\frac{d^{2}}{dz^{2}} + z^{2}\frac{d}{dz}]}
  \end{equation}
  \n   where $\lambda$ is real parameter and $ {\color{red}i^{2} = -1}$.\\
  
 \n For $\displaystyle{z = -iy ; y \in [0, \infty [ \,\, \text{ and } \,\, u(y) = \varphi(-iy)}$ we have\\
  \begin{equation}
 \displaystyle{H_{\lambda} u(y) =  H_{0, \lambda}u(y) = \lambda [-y u^{''}(y) + y^{2}u^{'}(y)] }
 \end{equation}
 \n  where $u'(y)$ denotes the first derivative of $u(y)$ and $u''(y)$ is its second derivative.\\
  
   \n In this case we will consider the following integral operator:\\
  \begin{equation}
\displaystyle{K_{0,\lambda}\psi(-iy) = \int_{0}^{\infty}\mathcal{N}_{0 ,\lambda}(y, s)\psi(-is)ds}
\end{equation}
\n where \\
\begin{equation}
\displaystyle{\mathcal{N}_{0,\lambda}(y, s) = \frac{1}{\lambda s}e^{-\frac{s^{2}}{2} }\int_{0}^{min(y, s)}e^{\frac{u^{2}}{2}}du}
\end{equation}

\n {\color{red}$\rhd$} $\displaystyle{H_{0, \lambda}}$  is formally anti-adjoint.\\

\n {\color{red}$\rhd$} Let $\displaystyle{u_{n}(y) = \frac{y^{n}}{\sqrt{n!}}}$ then\\

\begin{equation}
\displaystyle{ H_{\lambda}u_{n}(y) = \lambda[ -\frac{n(n-1)y^{n-1}}{\sqrt{n!}} + \frac{n y^{n+1}}{\sqrt{n!}}]}
\end{equation}

\n {\color{red} ($\bullet_{2}$)} In 1998, we have given  in {\color{blue}[14]} the boundary conditions at infinity for a description of all maximal dissipative extensions in Bargmann space of the minimal Heun's operator $\displaystyle{ H_{I} = z \frac{d^{2}}{dz^{2}} + z^{2} \frac{d}{dz}}$. The characteristic functions of the dissipative extensions have computed and some completeness theorems have obtained for the system of generalized eigenvectors. It is well known that the restriction $H_{I}^{min}$ of the closure of $H_{I}$ on the polynomials set $\mathcal{P}$ is symmetric.\\

\n  But  the minimal domain $\displaystyle{ D(H_{I}^{min}) = \{ \varphi \in \mathbb{B} , \exists \, p_{n} \in \mathcal{P} , p_{n} \longrightarrow \varphi , \exists \, \psi \in \mathbb{B} ; H_{I}p_{n} \longrightarrow \psi \}}$ of $\displaystyle{ H_{I}}$ is different of its maximal domain $D(H_{I}) = \{ \varphi \in \mathbb{B} ; H_{I}\varphi \in \mathbb{B} \}$. \\

\n {\color{red} ($\bullet_{3}$)} It is also well known that $\displaystyle{ H_{I}}$ is chaotic operator in Devaney's sense  {\color{blue}[8]}  (see {\color{blue}[7]}  or the reference {\color{blue}[16]} which used the results of {\color{blue}[15]}). In particular its spectrum is $\sigma(H_{I}) = \mathbb{C}$.\\

\n It follows that $\displaystyle{H_{0,\lambda} = i\lambda H_{I}}$ and $\displaystyle{H_{0,\lambda}^{min} = i\lambda H_{I}^{min}}$. $ H_{I}^{min}$ is given by action on the standard orthonormal basis $\displaystyle{\{e_{n}\}_{n\geq 0}}$ in  the Bargmann space $\mathbb{B}$:\\
\begin{equation}
\displaystyle{ H_{I}^{min}e_{n} = b_{n}e_{n+1} + a_{n}e_{n} + b_{n-1}e_{n-1}; a_{n} = 0, b_{n} = n\sqrt{n+1} > 0 , n \geq 1}
\end{equation}
Then $H_{I}^{min}$ can be represented in $\displaystyle{\mathbb{B}_{0} = \{\varphi \in \mathbb{B} ; \varphi(0) = 0\}}$ by an infinite tridiagonal matrix \\

\begin{equation}
\n \mathbb{H}_{I}^{min}= \left(
  \begin{array}{ c c c c c c c c  } 
     0 & \sqrt{2}& 0 & 0&\dots&\dots& \\ 
     \sqrt{2}  & 0 &2\sqrt{3} &0 & \dots&\dots&\\
    0 & 2\sqrt{3} & 0 & 3\sqrt{4}& \dots &\dots&\\
     0 & 0&3\sqrt{4} &0 & \ddots  & \dots& \\
     \vdots &\vdots & \vdots & \ddots & \ddots  & \ddots &\\
      \vdots & \vdots & \vdots & \vdots & \ddots &\ddots& \\
  \end{array} \right).
  \end{equation}
known as the Jacobi-Gribov matrix.\\
\begin{definition}
\n Let $\mathcal{H}$ be a separable Hilbert space and denote by $< ,  >$ the inner product in this space.\\
\n Let $T$ be a closed symmetric operator densely defined in  $\mathcal{H}$ , i.e. $T  \subset T^{*}$, with domain $D(T) \subset  \mathcal{H}$.\\

\n {\color{red}$\bullet_{1}$} The deficiency indices $n_{+}(T), n_{-}(T)$ are defined as follows:\\

\n $n_{+}(T)$ is the dimension of $\displaystyle{\mathfrak{M}_{z} = (T - zI)D(A); \Im m z \neq 0}$ and  $n_{-}(T)$ is the dimension of eigenspace $\displaystyle{\mathfrak{N}_{ \overline{z}}}$ of  $T$ corresponding to the eigenvalue $\overline{z}$ of the operator $T$\\
\n i.e.\\
\begin{equation}
\displaystyle{n_{\pm}(T) := dim(\mathcal{H} \ominus rang(T - zI)) = dim ker(T^{*} - \overline{z}I), \quad z \in \mathbb{C}_{\pm}}.
\end{equation} 

\n {\color{red}$\bullet_{2}$} A closed operator $T$ defined in $\mathcal{H}$ is said to be {\color{red}completely non-selfadjoin} if there is {\color{red}no subspace} reducing  $\mathfrak{B}$ of $\mathcal{H}$ such that the part of $T$ in this subspace is self-adjoint.\\

\n A completely non-selfadjoint symmetric operator is often referred to as {\color{red} simple}.\\
\end{definition}
\begin{remark}
\n {\color{red}$\bullet_{1}$} The theory of deficiency indices of closed symmetry operator in a complex Hilbert space is well-known and well-studied {\color{blue}[28]}.\\
 
\n {\color{red}$\bullet_{2}$}  Deficiency indices measure how far a symmetric operator is from being self-adjoint. Determining whether or not a symmetric operator is self-adjoint is important in physical applications because different self-adjoint extensions of the same operator yield different descriptions of the same system under consideration  {\color{blue}[28]}.\\

\n The deficiency indices $(n_{+}(T), n_{-}(T))$ of a closed symmetric operator $T$ in a complex Hilbert space $\mathcal{H}$ are also defined by\\
\begin{equation}
\displaystyle{n_{{\color{red}\pm}}(T) := dim\, ker(T^{*} \, {\color{red}\mp}\, iI)}
\end{equation}
\end{remark}
\begin{definition} [{\color{red} completely indeterminate case}]  (Krein {\color{blue}[24]})\\

\n A closed symmetric operator $T$ in a complex Hilbert space $\mathcal{H}$ is said {\color{red}completely indeterminate} if $n_{+}(T) = n_{-}(T) \neq 0$\\
\end{definition}
\n {\color{red}$\bullet_{3}$}  Let $\ell_{2}(\mathbb{N})$ be the Hilbert space of infinite sequences $\varphi = (\varphi_{1}, ...., \varphi_{n}, ...)$ with the inner product $\displaystyle{ < \varphi , \psi > =}$ $\displaystyle{ \sum_{n=1}^{\infty} }$ $\displaystyle{\varphi_{n}\overline{\psi}_{n}}$.\\

\n The matrix\\
 \begin{equation}
\mathcal{J} = \left(
  \begin{array}{ c c c c c c c c  } 
     a_{1} & b_{1}& 0 & 0&\dots&\dots& \\ 
     b_{1}  & a_{2} &b_{2} &0 & \dots&\dots&\\
    0 & b_{2} & a_{3} & b_{3}& \dots &\dots&\\
    0 & 0&b_{3} &a_{4} & \ddots  & \dots& \\
     \vdots &\vdots & \vdots & \ddots & \ddots  & \ddots &\\
      \vdots & \vdots & \vdots & \vdots & \ddots &\ddots& \\
  \end{array} \right).
  \end{equation}
\n  defines a symmetric operator $\mathfrak{T}$ on  $\ell_{2}(\mathbb{N})$ according to the formula\\
 \begin{equation}
 \left \{ \begin{array} {c} (\mathfrak{T}\varphi)_{n} = b_{n-1}\varphi_{n-1} + a_{n}\varphi_{n} + b_{n}\varphi_{n+1}, \quad n=1, 2, ..\\
 \quad\\
 \varphi_{0} = 0 \quad \quad \quad \quad \quad \quad  \quad \quad \quad \quad \quad \quad \quad \quad \quad \quad \quad \quad  \\
 \end{array} \right.
 \end{equation}
\n The closure $\mathfrak{T}^{min}$ with domain $D(\mathfrak{T}^{min})$ of the operator $\mathfrak{T}$ is the minimal closed symmetric operator generated by above expression.\\

\n To matrix $\mathcal{J}$ we assign the second-order difference equation:\\
\begin{equation}
\displaystyle{b_{n-1}\varphi_{n-1} + a_{n}\varphi_{n} + b_{n}\varphi_{n+1} = z\varphi_{n}, \quad z \in \mathbb{C},  n =2, 3, ..}
\end{equation}
\n It has two linearly independent polynomial solutions $\displaystyle{P(z) = (P_{n}(z))_{n=1}^{\infty}}$ and $\displaystyle{Q(z) = (Q_{n}(z))_{n=1}^{\infty}}$ :\\

\begin{equation}
\left \{ \begin{array} {c} \displaystyle{b_{n-1}P_{n-1} + a_{n}P_{n} + b_{n}P_{n+1} = zP_{n}, \quad z \in \mathbb{C}, \,\,  n =2, 3, ..}\\
\quad\\
\text{with the initial conditions}:  \quad \quad\quad \quad\quad \quad\quad \quad\quad \quad\quad \quad\\
\quad\\
P_{1}(z) =1 \quad \quad\quad \quad  \quad\quad \quad  \quad\quad \quad  \quad\quad \quad  \quad\quad \quad  \quad\quad \quad\\
\quad\\
\displaystyle{P_{2}(z) = \frac{z- a_{1}}{b_{1}}} \quad \quad\quad \quad\quad \quad  \quad\quad \quad  \quad\quad \quad\quad \quad  \quad\quad \quad\\

\end{array} \right.
\end{equation}

\n According to Berezanski's theory {\color{blue}[4] } on the second-order difference equations, $\displaystyle{(P_{n}(z))_{n=1}^{\infty}}$ will be called a sequence of polynomials of the first kind associated to $\mathcal{J}$.\\

\begin{equation}
\left \{ \begin{array} {c} \displaystyle{b_{n-1}Q_{n-1} + a_{n}Q_{n} + b_{n}Q_{n+1} = zQ_{n}, \quad z \in \mathbb{C}, \,\,  n =2, 3, ..}\\
\quad\\
\text{with the initial conditions}:  \quad \quad\quad \quad\quad \quad\quad \quad\quad \quad\quad \quad\\
\quad\\
Q_{1}(z) = 0 \quad \quad\quad \quad  \quad\quad \quad  \quad\quad \quad  \quad\quad \quad  \quad\quad \quad  \quad\quad \quad\\
\quad\\
\displaystyle{Q_{2}(z) = \frac{1}{b_{1}}} \quad \quad \quad\quad \quad\quad \quad  \quad\quad \quad  \quad\quad \quad\quad \quad  \quad\quad \quad\\

\end{array} \right.
\end{equation}

\n Still following Berzanskii, the sequence $\displaystyle{(Q_{n}(z))_{n=1}^{\infty}}$ is called a sequence of polynomials of the second kind associated  to $\mathcal{J}$ and to Kostyuchenko-Mirsoev {\color{red} [21]}, we can recall  the following Hellenger's theorem and the theorem of  Kostyuchenko-Mirsoev which are essential for further consideration :\\
 
 \begin{theorem} (Hellinger {\color{blue}[9]})\\
 
 \n  Suppose that there exists a point $z_{0} \in \mathbb{C}$ such that every solution $\displaystyle{\varphi = \varphi(z) = (\varphi(z))_{n=1}^{\infty}}$ of the second order difference equation\\
 
 \begin{equation}
 \displaystyle{b_{n-1}\varphi_{n-1} + a_{n}\varphi_{n} + b_{n+1}\varphi_{n+1} = z\varphi_{n} , n \geq 2}
 \end{equation}
 $$ \displaystyle{z, b_{n}, a_{n} \in \mathbb{C}, \quad b_{n} \neq 0}$$
 
 \n for $z = z_{0}$ satisfies the condition $\displaystyle{\sum_{n=1}^{\infty} \vert \varphi_{n}(z_{0}) \vert^{2} < \infty}$. Then for every solution $\varphi$ and $M > 0$ the series $\displaystyle{\sum_{n=1}^{\infty} \vert \varphi_{n}(z) \vert^{2}}$ converges uniformly in the disc $\displaystyle{\{ z ; \vert z - z_{0} \vert < M\}}$.\\
 \end{theorem}
 
 \n (A generalization of this theorem for difference equations of an arbitrary order and spaces $\displaystyle{\ell_{p} ;  1 \leq p \leq \infty}$ is contained in {\color{blue}[27]}).\\
 
 \n {\color{red}$\rhd_{4}$} Denote by $\ell_{m}^{2}$ the Hilbert space of quadratic summable sequences $\displaystyle{\phi = (\phi_{1}, \phi_{2} , .., \phi_{j}, ..)}$ where the vector column $\displaystyle{\phi_{j} \in \mathbb{C}^{m}}$, with inner product $\displaystyle{ < \phi, \Xi > = \sum_{j=1}^{\infty}\phi_{j}\overline{\Xi}_{j}}$. Also, denote by $\mathcal{J}_{m}$ an infinite matrix with the entries
  from $M_{m}(\mathbb{C})$ (By $M_{m}(\mathbb{C})$, we mean the space of $m$ by $m$ matrices with complex entries) of the form\\
 \begin{equation}
\mathcal{J}_{m} = \left(
  \begin{array}{ c c c c c c c c  } 
     A_{1} & B_{1}& 0 & 0&\dots&\dots& \\ 
     B_{1}^{*}  & A_{2} &B_{2} &0 & \dots&\dots&\\
    0 & B_{2}^{*} & A_{3} & B_{3}& \dots &\dots&\\
    0 & 0&B_{3}^{*} &A_{4} & \ddots  & \dots& \\
     \vdots &\vdots & \vdots & \ddots & \ddots  & \ddots &\\
      \vdots & \vdots & \vdots & \vdots & \ddots &\ddots& \\
  \end{array} \right).
  \end{equation}
 \n \quad\\
 
  \n where $A_{j} = A_{j}^{*}, det(B_{j}) \neq 0; \, j = 1,2, ....$ and $0$ is a zero matrix of order $m$. \\
  
  \n We will identify the matrix $\mathcal{J}_{m}$ with the operator $\mathfrak{T}$ defined as closure of the operator acting on the dense set of finite vectors  from $\ell_{m}^{2}$, where the action of this operator is described as following:\\
  
  \n {\color{red}$\circ_{1}$} In the {\color{red}vectors} $\displaystyle{\phi_{n} \in \mathbb{C}^{m}}$\\
  \begin{equation} 
  \displaystyle{(\mathfrak{T}\phi)_{n} = B_{n-1}^{*}\phi_{n-1} + A_{n}\phi_{n} + B_{n}\phi_{n+1} = z\phi_{n} \quad z \in \mathbb{C}, n \in \mathbb{N}}. \quad
  \end{equation}
  
  \n {\color{red}$\circ_{2}$} In the {\color{red}matrices} $\displaystyle{\Xi_{n} \in M_{m}( \mathbb{C})}$\\
  \begin{equation}
  \displaystyle{(\mathfrak{T}\Xi)_{n} = B_{n-1}^{*}\Xi_{n-1} + A_{n}\Xi_{n} + B_{n}\Xi_{n+1} = z\Xi_{n} \quad z \in \mathbb{C}, n \in \mathbb{N}}. \quad
  \end{equation}
\n  As  in scalar case, we can consider the matrix polynomial  solutions \\
  
  \n  $\displaystyle{P(z) = (P_{n}(z))_{n=1}^{\infty}}$ and $\displaystyle{Q(z) = (Q_{n}(z))_{n=1}^{\infty}}$  with initial conditions\\
$$ \displaystyle{P_{1}(z) = I, P_{2}(z) = B_{1}^{1}(z I - A_{1}); \quad Q_{1}(z) = 0, Q_{2}(z) = B_{1}^{-1}}$$

\begin{definition} ({\color{red} Completely indeterminate case}  of $\mathcal{J}_{m}$ {\color{blue}[25]}) \\

\n The completely indeterminate case of $\mathcal{J}_{m}$ holds if its defect indices are $(m, m)$.\\
\end{definition}
\begin{remark}
\quad \\

\n The defect indices of $\mathcal{J}_{m}$ satisfy the inequalities $0 \leq n_{+}(\mathcal{J}_{m}), n_{-}(\mathcal{J}_{m}), \leq m$ and they are not necessarily equal to each other.\\
\end{remark}
  \n In 1998, it was shown by A. Kostyuchenko and K. Mirsoev the following theorem: \\
  
 \begin{theorem} (Kostyuchenko-Mirsoev {\color{blue}[21]})\\
 
 \n The completely indeterminate case holds for the operator $\mathfrak{T}$ if and only if all solutions of the  equation with the boundary condition $\varphi_{0} = 0$:\\
 \begin{equation}
 \displaystyle{ (\mathfrak{T}\varphi)_{n} = z\varphi_{n}, \quad n = 1, 2, ...}\,\, \text{belongs to}\,\,  \ell_{m}^{2}(\mathbb{N}) \,\,\text{for} \,\, z = 0.\\
 \end{equation}
 \end{theorem}
 
 \n In 2001, a number of sufficient conditions  for complete indeterminacy in terms of the entries of $\mathcal{J}_{m}$ was obtained  by A. Kostyuchenko and K. Mirsoev  in {\color{blue}[22]}. We will applied these results to the following operator:\\
 
 \begin{equation}
 \displaystyle{H^{p,m} = A^{*p}(A^{m} + A^{*m})A^{p} \quad p, m = 1,2, .....}
 \end{equation}

 \n According to Berezanskii, Chap VII, 2 {\color{blue}[4]} in scalar case, it is well known that the deficiency numbers $n_{+}(\mathfrak{T}^{min})$ and $n_{-}(\mathfrak{T}^{min})$ of the operator $\mathfrak{T}^{min}$  satisfy the inequalities $0 \leq n_{+}(\mathfrak{T}^{min}) \leq 1$ and $0 \leq n_{-}(\mathfrak{T}^{min}) \leq 1$.\\

\n According the theorem 1.5 chap. VII of {\color{blue}[4]} and the results given in {\color{blue}[14]}, The main purpose of the present work is to give  in following section some new spectral properties of $H_{I}^{min}$ and in second section, we apply the results of A. Kostyuchenko and K. Mirsoev  to study the deficiency numbers of the generalized Heun's operator\\

$$ \displaystyle{H^{p,m} = A^{*p}(A^{m} + A^{*m})A^{p} \quad p, m = 1,2, .....}$$

\n acting on Bargmann space. In particular, here we find some conditions on the parameters $p$ and $m$ for that $ \displaystyle{H^{p,m}}$ to be completely indeterminate. It follows from these conditions that $ \displaystyle{H^{p,m}}$ is entire of the type minimal.\\
\n In section 3, we present some spectral properties of  integral operator $K_{0,\lambda}$ associated to $H_{0,\lambda} = i\lambda H_{I}$ on negative imaginary axis.\\

\section{{\color{red} New spectral properties of $H_{\lambda} = i\lambda A^{*}(A + A^{*})A$}}

\n Let $\displaystyle{H_{\lambda} = H_{0,\lambda} = i\lambda A^{*}(A + A^{*})A}$ with domaine $\displaystyle{D(H_{\lambda}) = \{\varphi \in \mathbb{B}_{0}; H_{\lambda}\varphi \in \mathbb{B}_{0}\}}$.\\

\n Now, we denote $\displaystyle{H_{\lambda_{\vert_{\mathcal{P}}}}}$ the above operator if we limit its domain to polynomials $\mathcal{P}_{0}$ where $\displaystyle{\mathcal{P}_{0} = \{p \in \mathcal{P} ;  p(0) = 0\}}$  and we denote the closure of this restriction by  $H_{\lambda}^{min}$. \\

\n Hence $H_{\lambda}$  is obviously an extension to this closure. \\ 

 \n  {\color{red} $\rhd$} It is well known that the minimal domain of $H_{\lambda}$  is different of its maximal domain contrary to minimal domain and maximal domain of $H_{\mu, \lambda}$ which coincide for $\mu \neq 0$.\\
 
 \n According Askey and Wilson in {\color{blue}[2]} on some hypergeometric orthogonal polynomials we deduce that  for above Jacobi matrix associated to $H_{I}$ that it is related the set of polynomials $P_{n}(x)$ of degree $n$ satisfying the recurrent relation,\\
 
 \begin{equation}
  \displaystyle{b_{n}P_{n+1}(x) + a_{n}P_{n}(x) + b_{n-1}P_{n-1}(x); = xP_{n}(x)}, n \geq 2
\end{equation}
\n with the following initial conditions:\\
 \begin{equation}
 \displaystyle{P_{0}(x) = 0 \quad P_{1}(x) = 1}
 \end{equation}
\quad\\
\n where  $\displaystyle{a_{n} = 0, \,\, \text{and}\,\, b_{n} = n\sqrt{n+1} }$.\\

\begin{definition}

\n The polynomials $P_{n}(x)$, which solve to above recurrence relation and subject to the initial conditions  $\displaystyle{P_{1}(x) = 1\,\, \text{and}\, \, P_{0}(x) = 0}$, are called polynomials of the first kind.\\

\n The polynomials $Q_{n}(x)$, which solve to above recurrence relation and subject to the initial conditions $\displaystyle{Q_{1}(x) = 1 \quad Q_{2}(x) = \frac{1}{\sqrt{2}}}$, are called polynomials of the second kind.\\

\begin{remark} ({\color{red} discrete Wronskian})\\

\n Let $\displaystyle{b_{n}u_{n+1} + b_{n-1}u_{n-1} = \lambda  u_{n} , n \geq 2}$ the spectral equation for $\mathcal{J}$. Then for every two solutions $u = (u_{n})$ and $v = (v_{n})$ of above equation with the same parameter $\lambda$ the discrete Wronskian defined by \\
\begin{equation}
\displaystyle{W(u, v) = b_{n}(u_{n}v_{n+1} - u_{n+1}v_{n})}
\end{equation}
 \n is independent of $n$\\
 
\n For the polynomials $(P_{n}(x))$ and $(Q_{n}(x))$ the wronskian is equal to one  for every $x$. It follows that \\

 \begin{equation}
\displaystyle{P_{n}Q_{n+1} - P_{n+1}Q_{n} = \frac{1}{b_{n}}  }
\end{equation}
\end{remark}

 \begin{definition} (Borzov {\color{blue}[6]})\\
 
 \n A polynomial set $\displaystyle{\{\psi_{n}(x)\}_{n=1}^{\infty}}$ is called a {\color{red}canonical polynomial system}  if it is defined by the following recurrence relations:\\
\begin{equation}
\displaystyle{ c_{n-1} \psi_{n-1}(x) + c_{n}\psi_{n+1} = x \psi_{n} \quad n \geq 1,\quad c_{0} = 0}
\end{equation}
\begin{equation}
\displaystyle{ \psi_{1}(x) = 1}
\end{equation}
 \end{definition}
\n  where the {\color{red}positive} sequence $\displaystyle{\{c_{n}\}_{n=1}^{\infty}}$ is given.\\
\end{definition}
 \begin{lemma}
 \n (i)  The polynomial set $\displaystyle{\{P_{n}(x)\}_{n=1}^{\infty}}$ is {\color{red}canonical polynomial system} \\
 
 \n (ii)  The polynomials $P_{n}(x)$ have real coefficients and fulfill the following parity conditions \\
 \begin{equation}
  \displaystyle{P_{n}(-x) = (-1)^{n-1}P_{n}(x)}.\\
  \end{equation}
  
   \begin{equation}
 \n (iii)  \quad \quad \quad \quad \quad \quad \quad \quad  \quad \quad   \displaystyle{b_{n-1}b_{n+1} \leq b_{n}^{2}} \quad \quad \quad \quad \quad \quad \quad \quad \quad \quad \quad \quad \quad \quad \quad \quad 
  \end{equation}
   \begin{equation}
 \n (iv) \quad \ \quad \quad \quad \quad \quad \quad \quad \quad \quad \displaystyle{\sum_{n=1}^{\infty} \frac{1}{b_{n}} < \infty}\quad \quad \quad \quad \quad \quad \quad \quad    \quad \quad \quad \quad \quad \,\,\,\,
  \end{equation}
 \n (v)  The operator  $H_{\lambda}^{min}$ has the deficiency indices $(1, 1)$.\\
 \end{lemma} 
 \n {\color{red}{\bf Proof}}\\
 
 \n (i) As the sequence $b_{n}$ is positive then the proof of this property is trivial by taking $c_{n} = b_{n}$.\\
 
 \n (ii) As the coefficients of  $\displaystyle{P_{1}(x) = 1}$ and the coefficients of  $\displaystyle{P_{2}(x) =  \frac{x}{b_{1}}}$ are real then we deduce by recurrence that The polynomials $P_{n}(x)$ have real coefficients. Now as $\displaystyle{P_{2}(-x) =  -\frac{x}{b_{1}} = (-1)P_{2}(x)}$  then if we suppose that  $ \displaystyle{P_{n}(-x) = (-1)^{n-1}P_{n}(x)}$ and   $\displaystyle{P_{n-1}(-x) = (-1)^{n-2}P_{n-1}(x)}$ then from the recurrence relation \\
$$ \displaystyle{b_{n}P_{n+1}(-x)  = -xP_{n}(-x)  - b_{n-1}P_{n-1}(-x) }$$
\n we deduce that\\
$$ \displaystyle{b_{n}P_{n+1}(-x)  = - x(-1)^{n-1}P_{n}(x)  - b_{n-1}(- 1)^{n-2}P_{n-1}(x)}$$
 \n and \\
 $$ \displaystyle{b_{n}P_{n+1}(-x)  = (-1)^{n}[xP_{n}(x)  - b_{n-1}P_{n-1}(x) ] = (-1)^{n}b_{n}P_{n+1}(x)}$$
  \n It follows that \\
  $$ \displaystyle{P_{n+1}(-x)  = (-1)^{n}P_{n+1}(x)}.$$
  \n (iii) Let $b_{n} = n\sqrt{n+1}$ then $b_{n-1} = (n-1)\sqrt{n}$ and $b_{n+1} = (n+1)\sqrt{n+2}$. It follows that\\
  \n {\color{red} $\rhd_{1}$} $\displaystyle{b_{n-1}b_{n+1} = (n-1)(n+1)\sqrt{n(n + 2)}}$ and $\displaystyle{b_{n}^{2} = n^{2}(n+1)}$. This implies that\\
  
  \n $\displaystyle{b_{n-1}b_{n+1} \leq b_{n}^{2} \iff (n-1)^{2} \leq n^{3} \iff 3n \geq 2}$.\\
  
\n {\color{red} $\rhd_{2}$}  As  $3n \geq 2 $ holds for all $n \geq 1$, we deduce that $\displaystyle{b_{n-1}b_{n+1} \leq b_{n}^{2}}$.\\

\n (iv) It is well known that the following series $\displaystyle{\zeta (\alpha) = \sum_{n= 1}^{\infty} \frac{1}{n^{\alpha}}}$ converges for $\alpha >1$, where $\alpha$ is real, then as  $\displaystyle{ \sum_{n=1}^{\infty}\frac{1}{b_{n}} =  \sum_{n=1}^{\infty}\frac{1}{n\sqrt{n+1}} \equiv  \sum_{n=1}^{\infty}\frac{1}{n^{\frac{3}{2}}}}$, it follows  that  $\displaystyle{ \sum_{n=1}^{\infty}\frac{1}{b_{n}} < \infty}$.\\

\n (v) By using theorem 1.5, Ch. VII {\color{blue}[4]}.\\

\n For any operator $T$ by $\sigma(T)$, $\sigma_{ess}(T)$ and $\sigma_{p}(T)$ we denote the spectrum, the essential spectrum and the point spectrum of $T$, respectively. \\

\begin{proposition}

\n Let $\displaystyle{H_{I} = z\frac{d^{2}}{dz^{2}} + z^{2}\frac{d}{dz}}$ acting on Bargmann space $\mathbb{B}_{0}$. For all $\displaystyle{ \varphi(z) =  \sum_{n=1}^{n} \varphi_{n} \frac{z^{n}}{\sqrt{n!}}}$, $H_{I}$ can be defined by\\
\begin{equation}
\left \{ \begin{array} {c}\displaystyle{[H_{I}\varphi](z) = \sum_{n=1}^{\infty} (H_{I})_{n} \frac{z^{n}}{\sqrt{n!}}} \quad  \quad \quad \quad \quad \quad \quad\\
\quad\\
\text{where} \quad \quad \quad \quad \quad \quad \quad \quad \quad  \quad \quad \quad  \quad \quad \quad \\
\quad\\
(H_{I}\varphi)_{1} = \sqrt{2} \varphi_{2} \quad \quad \quad \quad \quad \quad\ \quad \quad \quad  \quad \quad  \\
\quad\\
(H_{I}\varphi)_{n} = (n-1)\sqrt{n} \varphi_{n-1} + n\sqrt{n + 1} \varphi_{n+1}\\
\end{array} \right.
\end{equation}

\n Then the point spectrum of  $H_{I}$ is $\mathbb{C}$.\\
\end{proposition}
\n {\color{red}{\bf Direct proof}}\\

\n This direct proof is based on the following classical proposition:\\

\begin{proposition}
The Bertrand series $\displaystyle{\sum_{n > 1} u_{n}}$ with general term $\displaystyle{u_{n} = \frac{1}{n ln(n)^{\beta}}}$ converges if and only if $\beta > 1$.
\end{proposition}

\n Now, let's consider the sequence  $\displaystyle{(u_{n}(\xi)), n = 1,2, ...}; \xi \in \mathbb{C}$ defined by the recurrence relation:\\

$ \left \{ \begin{array} {c} \displaystyle{ u_{1}(\xi) = 1}  \quad \quad \quad \quad   \quad \quad \quad   \quad \quad \quad   \quad \quad \quad  \quad \quad \quad  \quad \quad \quad \\
\quad\\
\displaystyle{u_{2}(\xi) = \frac{\xi}{\sqrt{2}}} \quad \quad \quad   \quad \quad \quad   \quad \quad \quad  \quad \quad \quad   \quad \quad \quad  \quad \quad \quad  \\
\quad\\
\displaystyle{ (n-1)\sqrt{n} u_{n-1}(\xi) + n\sqrt{n + 1} u_{n+1}(\xi) = \xi u_{n}(\xi)} \quad n \geq 2\\
\end{array} \right. $ \hfill { } {\color{blue} ($\star$)}\\

\n and let $\displaystyle{\varphi_{\xi}(z) = \sum_{n=1}^{\infty}u_{n}(\xi)\frac{z^{n}}{\sqrt{n!}}}$, where $\displaystyle{u_{n}(\xi)}$ are defined by {\color{blue} ($\star$)}.\\

\n {\color{red}$\rhd_{1}$} It is clear that $\displaystyle{H_{I}\varphi_{\xi} = \xi \varphi_{\xi}} $.\\

\n {\color{red}$\rhd_{2}$} For that $\varphi_{\xi}$ belongs to $\mathbb{B}_{0}$ we must to prove that for all $\xi \in \mathbb{C}$ we have $\displaystyle{(u_{n}(\xi)) \in \ell_{2}(\mathbb{N})}$. For this, we will shown that \\
$$\exists \, \, n_{0} \, \text{such that} \,\,\forall \,\, n \geq n_{0}\, \text{ we have} \,\, \displaystyle{\vert u_{n}(\xi) \vert \leq \frac{M}{\sqrt{n}ln(n)}} \quad \quad \quad \quad {\color{blue} (\star\star)}$$ 
\n where $M$ is a constant not dependent of $n$.\\

\n To determine $n_{0}$ we use the below relation when $n \longrightarrow \infty$\\
\begin{equation}
\displaystyle{\frac{1}{2n^{\frac{3}{2}}ln(n)} \sim \frac{1}{\sqrt{n+1}ln(n + 1)} - \frac{\sqrt{n-1}}{\sqrt{n(n -1)}ln(n -1)} - \frac{\vert \xi\vert}{n\sqrt{n(n+1)}ln(n)}}
\end{equation}
which proves that the second part of the above equality is positive for $n \geq n_{0}$.\\ 

\n Now, we set $\displaystyle{ M = max\{\vert u_{n_{0}}\vert ln(n_{0}), \vert u_{n_{0}-1}\vert ln(n_{0}-1)\}}$ and we deduce {\color{blue} ($\star\star$)} by recurrence. We use the following inequalities \\
$$\displaystyle{\vert u_{n+1} \vert \leq \vert \xi \vert \frac{\vert u_{n} \vert}{n \sqrt{n+1}}+ \frac{(n-1)\sqrt{n}\vert u_{n-1} \vert}{n\sqrt{n+1}}}$$
\n and\\
 $$\displaystyle{ \frac{(n-1)\sqrt{n}}{n\sqrt{(n+1)(n -1)}ln(n-1)} + \frac{\vert \xi \vert}{n\sqrt{n(n+1)}ln(n)} \leq \frac{1}{\sqrt{(n+1)}ln(n+1)}}$$
 \n It follows that \\ 
 $$\displaystyle{\vert u_{n}(\xi) \vert^{2} \leq \frac{M^{2}}{nln^{2}(n)} \,\, \text{and}\,\, \sum_{n=1}^{\infty}\vert u_{n}(\xi) \vert^{2} \leq M^{2}} \sum_{n=1}^{\infty}{\frac{1}{nln^{2}(n)} < \infty}$$
 \n This implies that  $\displaystyle{(u_{n}(\xi)) \in \ell_{2}(\mathbb{N})}$ for all $\xi \in \mathbb{C}$.\\
 
 \n {\color{red}{\bf Second proof}} \\
 
 \n The second proof is based on the following classical lemma:\\

\begin{lemma} ({\color{red}Raabe-Duhamel test})\\

 \n We suppose, $\displaystyle{\forall \, n > 0  \, \, a_{n} > 0}$ \\
\n {\color{red}$\rhd_{1}$} If \\
 $$\displaystyle{\exists \,\, \alpha \in \mathbb{R}, \frac{a_{n+1}}{a_{n}} = 1 - \frac{\alpha}{n} + o(\frac{1}{n}),}$$
 \n then\\
 
 \n (i) $\displaystyle{\alpha > 1 \Longrightarrow \sum a_{n} }$ converges\\
 
  \n (ii) $\displaystyle{\alpha  < 1 \Longrightarrow \sum a_{n} }$ diverges\\
  
  \n {\color{red}$\rhd_{2}$} Same conclusions if \\  
   $$\displaystyle{\exists \,\, \alpha \in \mathbb{R}, \frac{a_{n+1}}{a_{n}} = 1 - \frac{1}{n} - \frac{\alpha}{n ln(n)} + o(\frac{1}{nln(n)}),}$$
\end{lemma}
 
\n and  the theorem 2.3 of the reference {\color{blue}[16]}\\

\begin{corollary}
$\sigma_{ess}(H_{I}) = \emptyset $ .
\end{corollary}

\section{{\color{red} Some spectral properties of  Integral operator $K_{0,\lambda}$ a right inverse  of $H_{0,\lambda}$ on negative imaginary axis}}

\n For $\mu > 0$, let $\sigma(\mu)$ be the smallest eigenvalue of  the operator:\\
\begin{equation}
\displaystyle{H_{\mu,\lambda} = \mu A^{*}A + i\lambda A^{*}(A + A^{*} )A}
\end{equation}
in the orthogonal complement of the vacuum where $A^{*}$ and $A$ are the creation and annihilation operators.\\

\n it is well known that $\sigma(\mu)$  extends to a positive, increasing, analytic function on the whole real line and that the limit value $\sigma(0)$ is an eigenvalue of $\displaystyle{H_{\lambda} := H_{0,\lambda} = i\lambda A^{*}(A + A^{*} )A}$ in particular, $\sigma(0) \neq 0$.\\

\n Despite the difficulty of the {\color{red}absence} of any relation between the domains of the self-adjoint and anti-adjoint parts of $H_{\mu,\lambda}$, this operator  has a fine spectral property in the Bargmann representation, if we restrict to an imaginary semi-axis, its inverse is an integral operator with a positive kernel, which allows us to apply the Krein-Rutman theorem {\color{blue}[26]} and the Jentzsch theorem {\color{blue}[19]}.\\

\n {\color{red}$\bullet$} If $\displaystyle{e_{n}(z) = \frac{z^{n}}{\sqrt{n!}}; z = x + iy}$ is the usual basis of Bargmann space then \\$\displaystyle{\tilde{e}_{n}(z) = \frac{(iz)^{n}}{\sqrt{n!}}; z = x + iy}$ is also orthonormal basis of Bargmann space.\\

\n Now, if we restrict to an imaginary semi-axis for example $z = -iy , y \in [0, +\infty[$, we deduce that\\

\n $\displaystyle{e_{n}(-iy) = \frac{(-iy)^{n}}{\sqrt{n!}}}$ and $\displaystyle{\tilde{e}_{n}(-iy) = \frac{(i(-iy))^{n}}{\sqrt{n!}} = \frac{y^{n}}{\sqrt{n!}}: = u_{n}(y)}$.\\

\n We recall that on $\displaystyle{\mathbb{B}_{0} = \{\varphi \in \mathbb{B} ; \varphi(0) = 0\}}$, it was well known that an explicit inverse of $H_{\mu, \lambda}$ restricted on imaginary axis ; $y \in [0,  +\infty[$ is given by\\

$$\displaystyle{K_{\mu,\lambda}\psi(-iy):= H_{\mu, \lambda}^{-1} \psi(-iy) = \int_{0}^{\infty}\mathcal{N}_{\mu ,\lambda}(y, s)\psi(-is)ds}$$

\n where \\
$$\displaystyle{\mathcal{N}_{\mu,\lambda}(y, s) = \frac{1}{\lambda s}e^{-\frac{s^{2}}{2} - \frac{\mu}{\lambda}s}\int_{0}^{min(y, s)}e^{\frac{u^{2}}{2} +\frac{\mu}{\lambda}u}du}$$
\n It follows that\\
\begin{equation}
\displaystyle{K_{\mu,\lambda}\psi(-iy):= H_{\mu, \lambda}^{-1} \psi(-iy) = \int_{0}^{y}e^{\frac{u^{2}}{2} + \frac{\mu}{\lambda}u}du \int_{u}^{\infty}\frac{1}{\lambda s}e^{-\frac{s^{2}}{2} - \frac{\mu}{\lambda}s}ds}
\end{equation}

\n For $\mu = 0$,  we deduce that:\\

\n $\displaystyle{K_{0,\lambda}\psi(-iy) = \frac{1}{\lambda} \int_{0}^{y}e^{\frac{u^{2}}{2}} du \int_{u}^{\infty}\frac{1}{s}e^{-\frac{s^{2}}{2}} \psi(-is)ds}$ $\displaystyle{ = \frac{1}{\lambda} \int_{0}^{y} du \int_{u}^{\infty}\frac{1}{s}e^{-\frac{1}{2}(s^{2} -u^{2})} \psi(-is)ds}$

$$\displaystyle{= \frac{1}{\lambda} [-\int_{0}^{y} du\int_{0}^{u}\frac{1}{s}e^{-\frac{1}{2}(s^{2} -u^{2})} \psi(-is)ds + \int_{0}^{y}du \int_{0}^{\infty}\frac{1}{s}e^{-\frac{1}{2}(s^{2} -u^{2})} \psi(-is)ds}$$

\begin{remark}

\n {\color{red}$\bullet_{1}$} It well known that $K_{\mu,\lambda}$ is compact if $\mu \neq 0$ because $H_{\mu, \lambda}$ satisfies the following non trivial  properties see {\color{blue}[10]}\\

\n (i) $\displaystyle{ \rho(H_{\mu, \lambda}) \neq \emptyset}$ where $\rho(H_{\mu, \lambda})$ is the resolvent set of $H_{\mu, \lambda}$.\\

\n (ii)  $\displaystyle{ D_{min}(H_{\mu, \lambda}) =  D_{max}(H_{\mu, \lambda}) }$ where\\

 \n $\displaystyle{D_{max}(H_{\mu, \lambda}) = \{\varphi \in \mathbb{B}; H_{\mu, \lambda}\varphi \in \mathbb{B}\} }$.\\
  
\n  $\displaystyle{D_{min}(H_{\mu, \lambda}) = \{\varphi \in \mathbb{B}; \exists p_{n} \in \mathcal{P}, \exists \psi \in \mathbb{B}; \lim\limits_{n \longrightarrow +\infty}p_{n} = \varphi \, \text{and} \,\,  \lim\limits_{n \longrightarrow +\infty}H_{\mu, \lambda}p_{n} = \psi \}}$.\\

\n (iv)  For $\mu > 0$ $\displaystyle{\vert\vert H_{\mu, \lambda}\varphi \vert\vert \geq \mu \vert\vert A\varphi \vert\vert \, \forall \varphi \in D_{max}(H_{\mu, \lambda})}$.\\

\n (v) the injection of $\displaystyle{D_{max}(H_{\mu, \lambda}) }$ in $\displaystyle{D(A) = \{\varphi \in \mathbb{B} ; \frac{d}{dz}\varphi \in  \mathbb{B}\}}$ is continuous.\\

\n (vi)  the injection of $D(A)$ in $\mathbb{B}$ is compact.\\

\n {\color{red}$\bullet_{2}$} It well known that for $\mu = 0$, $H_{\mu, \lambda}$ satisfies some properties  which are to far form above properties:\\

\n (i) $\displaystyle{\rho(H_{0, \lambda}) = \emptyset}$ where $\rho(H_{0, \lambda})$ is the resolvent set of $H_{0, \lambda}$.\\

\n (ii)  $\displaystyle{D_{min}(H_{0, \lambda}) \neq   D_{max}(H_{0, \lambda}) }$.\\

 \n {\color{red}$\bullet_{3}$} The end of this section will be devoted to prove the compactness of $K_{0,\lambda}$ as a limit of finite rank operators. For this, we start recalling  some classical properties of compact operators and of operators of finite rank before to give the action of $K_{0,\lambda}$ on the basis $\tilde{e}_{n}(-iy)$.\\
\end{remark}

\begin{definition}

\n Let $\mathcal{H}$ be Hilbert space and $T$ be a bounded linear operator on $\mathcal{H}$. \\

\n (i) Let $ \Im mT = T( \mathcal{H})$, $T$ is said to be of rank $r$ ($r < + \infty$) if $dim \Im mT = r$. The class of operators of rank $r$ is denoted by $\mathcal{K}_{r}(\mathcal{H})$.\\
\n (ii) $T$ is said compact if, for any bounded sequence $(\varphi_{n})$ in $\mathcal{H}$, the sequence $(T\varphi_{n})$ contains a convergent subsequence. Equivalently, $T$ is compact when it maps the unit ball $\mathcal{B}$ in $\mathcal{H}$ to a pre-compact set in $\mathcal{H}$ ( A set in a topological space is pre-compact if its closure is compact).\\
\end{definition}
\begin{theorem}
$\displaystyle{T \in \mathcal{K}_{r}(\mathcal{H}) \iff T^{*} \in \mathcal{K}_{r}(\mathcal{H})}$
\end{theorem}

\n {\color{red}{\bf Proof}}\\

\n Let $\displaystyle{T \in \mathcal{K}_{r}(\mathcal{H})}$ and let $\displaystyle{\psi_{1}, \psi_{2}, ...., \psi_{r}}$  be an orthonormal basis in $\Im mT$. Then for any $\displaystyle{\varphi \in \mathbb{H}}$,  we have \\
$$\displaystyle{T\varphi = \sum_{n=1}^{r} < T\varphi , \psi_{n} >\psi_{n} = \sum_{n=1}^{r} < \varphi , T^{*}\psi_{n} >\psi_{n} }$$
\n Denote $\xi_{n} = T^{*}\psi_{n}$, then $\displaystyle{T\varphi = \sum_{n=1}^{r} < \varphi , \xi_{n} >\psi_{n}}$. Moreover\\
$$\displaystyle{< T\varphi, \psi > = \sum_{n=1}^{r} < < \varphi , \xi_{n} >\psi_{n}, \psi  > = \sum_{n=1}^{r} < \varphi , < \psi, \psi_{n} >\xi_{n}  = < \varphi, T^{*}\psi >}$$
\n Therefore $$\displaystyle{T^{*}\psi = \sum_{n=1}^{r} < \psi , \psi_{n} >\xi_{n}}$$
\n and thus $\displaystyle{T^{*} \in \mathcal{K}_{r}(\mathcal{H})}$\\

\begin{proposition}
An operator-norm limit of compact operators is compact.
\end{proposition}

\n {\color{red} {\bf Proof}}\\

\n Let $T_{n} \longrightarrow T$ in uniform operator norm, with compact $T_{n}$. 

\n Given $\epsilon  > 0$, let $n$ be sufficiently large such that $\displaystyle{\vert\vert T_{n} - T\vert\vert < \frac{\epsilon}{2}}$. Since $T_{n}(\mathcal{B})$ is pre-compact, there are finitely many $\psi_{1}, . . . , \psi_{r}$ such that for any $\varphi \in \mathcal{B}$ there is $i$ such that $\displaystyle{\vert\vert T_{n}\varphi - \psi_{i} \vert\vert < \frac{\epsilon}{2}}$. By the triangle inequality we deduce that \\
$$\displaystyle{\vert\vert T\varphi - \psi_{i} \vert\vert < \vert\vert T\varphi - T_{n}\varphi \vert\vert  +  \vert\vert T_{n}\varphi -   \psi_{i} \vert\vert  < \epsilon}$$
\n Thus, $T(\mathcal{B}) $ is covered by finitely many balls of radius $\epsilon$.\\

\begin{lemma}
Let $\mathcal{H}$ be Hilbert space and $T$ be a bounded operator on $\mathcal{H}$ then if $T$ has finite rank then $T$ is compact.
\end{lemma}

\n {\color{red} {\bf Proof}}\\

\n Since $T$ has finite rank, the space $ \Im mT = T( \mathcal{H})$ is a finite-dimensional normed space. Furthermore, for any bounded sequence $(\varphi_{n})$ in $\mathcal{H}$, the sequence $(T\varphi_{n})$ is bounded in $ \Im mT$ , so by the Bolzano-Weierstrass theorem this sequence must contain a convergent subsequence. Hence $T$ is
compact.\\

\begin{theorem}
\quad\\
\n A compact operator $T$ defined on Hilbert space $\mathcal{H}$ is an operator norm limit of finite rank operators.\\
\end{theorem}

\n {\color{red}{\bf Proof}}\\

\n Let $\mathcal{B}$ be the closed unit ball in $\mathcal{H}$. Since $T(\mathcal{B})$ is pre-compact it is totally bounded, so for given $\epsilon > 0$ cover $T(\mathcal{B})$ by open balls of radius $\epsilon$ centered at points $\psi_{1}, . . . , \psi_{r}$ . Let $P$ be the orthogonal projection to the finite-dimensional subspace $\mathcal{F}$ spanned by the $\psi_{i}$ and define $\displaystyle{ T_{\epsilon} = PoT }$. Note that for any $\psi \in \mathcal{H}$ and for any $\psi_{i}$\\
$$\displaystyle{\vert\vert P\psi - \psi_{i} \vert\vert \leq \vert\vert \psi - \psi_{i} \vert\vert}$$
\n since $\displaystyle{\psi = P\psi + \xi}$ with $xi$ orthogonal to  $\psi_{i}$. For $\varphi \in \mathcal{H}$ with $\vert\vert \varphi \vert\vert \leq 1$, by construction there is $\psi_{i}$ such that $\vert\vert T\varphi  - \psi_{i} \vert\vert < \epsilon $. Then\\
 $$\displaystyle{ \vert\vert T\varphi  - T_{\epsilon}\varphi \vert\vert  \leq \vert\vert T\varphi  - \psi_{i} \vert\vert + \vert\vert T_{\epsilon}\varphi  - \psi_{i} \vert\vert  < \epsilon + \epsilon}$$
\n Thus $\displaystyle{T_{\epsilon} \longrightarrow T}$ in operator norm as $\epsilon \longrightarrow 0$.\\ 

\begin{lemma} ({\color{red}Action of $K_{0,\lambda}$ on the basis $\tilde{e}_{n}(-iy)$})\\

\n Let $\displaystyle{\tilde{e}_{n}(-iy) = u_{n}(y) = \frac{y^{n}}{\sqrt{n!}}}$ then \\
\begin{equation}
\displaystyle{K_{0,\lambda}u_{1} : = v_{1}(y) = \frac{1}{\lambda} \int_{0}^{y} e^{\frac{u^{2}}{2}} du \int_{u}^{\infty} e^{-s^{2}}ds}
\end{equation}

\begin{equation}
\displaystyle{K_{0,\lambda}u_{n+1} = \frac{n-1}{\sqrt{n(n+1)}} K_{0,\lambda}u_{n-1} + \frac{u_{n}}{\lambda n\sqrt{n+1}}, n \geq 1} 
\end{equation}
\n i.e  \\
$$\quad\quad \quad \quad \quad \quad \quad \quad \displaystyle{v_{n+1} = \frac{n-1}{\sqrt{n(n+1)}} v_{n-1} + \frac{u_{n}}{\lambda n\sqrt{n+1}} \quad \quad \quad \quad \quad \quad \quad \quad(2.4)_{bis}}$$
\end{lemma}
\n {\color{red}{\bf Proof}}\\

\n Let $\displaystyle{v_{n+1}(y) = K_{0,\lambda}u_{n+1} = \frac{1}{\lambda} \int_{0}^{y}e^{\frac{u^{2}}{2}} du \int_{u}^{\infty}\frac{1}{s}e^{-\frac{s^{2}}{2}} \frac{s^{n+1}}{\sqrt{(n+1)!}}ds}$\\
$\displaystyle{ = \frac{1}{\lambda} \int_{0}^{y}e^{\frac{u^{2}}{2}} du \int_{u}^{\infty}  e^{-\frac{s^{2}}{2}} \frac{s^{n}}{\sqrt{(n+1)!}}ds}$ $\displaystyle{ =  \frac{1}{\lambda} \int_{0}^{y}e^{\frac{u^{2}}{2}} du \int_{u}^{\infty}  se^{-\frac{s^{2}}{2}} \frac{s^{n-1}}{\sqrt{(n+1)!}}}$\\

\n Setting $\displaystyle{v'(s) = se^{-\frac{s^{2}}{2}}}$ and $\displaystyle{u(s) = s^{n-1}}$, this implies that $\displaystyle{v(s) = - e^{-\frac{s^{2}}{2}}}$ and $\displaystyle{u'(s) = (n-1) s^{n-2}}$.\\

\n Now we use an integration by part to get:\\

\n $\displaystyle{ \int_{u}^{\infty}  e^{-\frac{s^{2}}{2}} s^{n}ds = [-e^{-\frac{s^{2}}{2}} s^{n-1}]_{u}^{\infty} + (n-1)\int_{u}^{\infty}  e^{-\frac{s^{2}}{2}} s^{n-2}ds}$.\\

\n $\displaystyle{= e^{-\frac{u^{2}}{2}} u^{n-1} + (n-1)\int_{u}^{\infty}  e^{-\frac{s^{2}}{2}} s^{n-2}ds}$.\\

\n It follows that \\

\n $\displaystyle{v_{n+1}(y) = \frac{1}{\lambda \sqrt{(n +1)!}}[ \int_{0}^{y}e^{\frac{u^{2}}{2}} e^{-\frac{u^{2}}{2}} u^{n-1}du + (n-1)\int_{0}^{y}e^{\frac{u^{2}}{2}}\int_{u}^{\infty}  e^{-\frac{s^{2}}{2}} s^{n-2}ds}$].\\

\n $\displaystyle{= \frac{1}{\lambda \sqrt{(n +1)!}} \frac{y^{n}}{n} + \frac{n-1}{\lambda \sqrt{(n +1)!}}\int_{u}^{\infty}  e^{-\frac{s^{2}}{2}} s^{n-2}ds}$.\\
\n $\displaystyle{= \frac{1}{\lambda \sqrt{(n +1)!}} \frac{\sqrt{n!} u_{n}}{n} + \frac{n-1}{\sqrt{n(n+1)}} \frac{1}{\lambda \sqrt{(n -1)!}}\int_{u}^{\infty}  e^{-\frac{s^{2}}{2}} s^{n-2}ds}$.\\

\n $\displaystyle{= \frac{u_{n}}{\lambda n\sqrt{(n +1)}}  + \frac{n-1}{\sqrt{n(n+1)}} v_{n-1}}$.\\

\begin{corollary}
\quad\\

\n (i) $K_{0,\lambda}$  is a right inverse of $H_{\lambda}$, i.e $H_{\lambda} K_{0,\lambda} = I$.\\

\n (ii) $H_{\lambda}$ is surjective.\\
\end{corollary}

\n {\color{red}{\bf Proof}}\\

\n (i) As $\displaystyle{u_{n} = \frac{y^{n}}{\sqrt{n!}}}$ then we deduce that \\
$$\displaystyle{ H_{\lambda}u_{n} = \lambda[ -\frac{n(n-1)y^{n-1}}{\sqrt{n!}} + \frac{n y^{n+1}}{\sqrt{n!}}]}$$
\n and\\
$$\displaystyle{ \frac{H_{\lambda}u_{n}}{n\sqrt{n + 1}} = -\frac{(n-1)u_{n-1}}{\sqrt{n(n +1)}} + u_{n+1}}$$
\n Now, from above recurrence relation :\\

\n $\displaystyle{K_{0,\lambda} u_{n+1} = \frac{u_{n}}{\lambda n\sqrt{(n +1)}}  + \frac{n-1}{\sqrt{n(n+1)}} K_{0,\lambda}u_{n-1}}$.\\

\n we deduce that\\

\n $\displaystyle{H_{\lambda}K_{0,\lambda} u_{n+1} = \frac{H_{\lambda} u_{n}}{\lambda n\sqrt{(n +1)}}  + \frac{n-1}{\sqrt{n(n+1)}} H_{\lambda}K_{0,\lambda}u_{n-1}}$.\\

\n $\displaystyle{ = -\frac{(n-1)}{\sqrt{n(n +1)}}u_{n-1} + u_{n+1} + \frac{n-1}{\sqrt{n(n+1)}} H_{\lambda}K_{0,\lambda}u_{n-1}}$.\\

\n Now as $K_{0,\lambda}u_{2}(y) = \frac{u_{1}(y)}{\lambda\sqrt{2}}$ then \\

\n $\displaystyle{H_{\lambda}K_{0,\lambda}u_{2}(y) = \frac{y^{2}}{\sqrt{2}} = u_{2}(y)}$.\\ 

\n and by recurrence  we deduce that\\

\n $\displaystyle{H_{\lambda}K_{0,\lambda} u_{n+1} = -\frac{(n-1)}{\sqrt{n(n +1)}}u_{n-1} + u_{n+1} + \frac{n-1}{\sqrt{n(n+1)}} u_{n-1} = u_{n+1} \, \, \forall \,\, n}$.\\

\n It follows by taking the range of $K_{0,\lambda}$ as domain of $H_{\lambda}$ that $\displaystyle{H_{\lambda}K_{0, \lambda} = I}$.\\

\n (ii) It well known that if an operator $T$ has a right inverse, then $T$ is surjective. Conversely, if $T$ is surjective and the axiom of choice is assumed, then $T$ has a right inverse (this assertion cannot be proved without the axiom of choice) see {\color{blue} [5]}.\\
\begin{remark}
\quad\\
\n (i) Let $\displaystyle{ \mathcal{P}(\mathbb{C}) }$ be the set of polynomials, and $\displaystyle{A = \frac{d}{dz} }$. We would like to undo differentiation, so we integrate:\\
$$\displaystyle{\mathbb{J}(p) = \int_{0}^{z} p(\xi)d\xi}$$
\n The fundamental theorem of calculus says that the derivative of this integral is $p$; that is, $\displaystyle{A\mathbb{J} = I_{\mathcal{P}(\mathbb{C})}}$. So $\mathbb{J}$ is a right inverse of $A$; it provides a solution not the only one  of the differential equation :\\
$$\displaystyle{\frac{d}{dz}q(z) = p(z)}$$
\n If we try things in the other direction, there is a problem:\\
$$\displaystyle{\mathbb{J}A(p) = \int_{0}^{z} p^{'}(\xi)d\xi = p(z) - p(0)}$$
\n That is, $\mathbb{J}A$ sends $p$ to $ p - p(0)$, which is not the same as $p$. So $\mathbb{J}$ is not a left inverse to $A$; since $A$ has a nonzero null space, we'll see that no left inverse can exist.\\

\n (ii) Consider the space $E$ of real sequences, the linear mapping $T$ that maps a sequence $(a_{0}, a_{1},.......)$ to the sequence $(0, a_{0}, a_{1},......)$ and the linear mapping $S$ that maps a sequence $(a_{0}, a_{1}, a_{2},......)$ to the sequence $(a_{1}, a_{2},......)$. It is clear that $ST= I$. Now consider the sequence $a = (1,0,0,0,......)$. We have $S(a) = 0$ where $0$ is the sequence that vanishes identically and also $TS(a)=0$ hence $TS \neq I$.

\end{remark}

\begin{remark}
\quad\\

\n {\color{red}$\rhd_{1}$} $v_{1}(y) $ is a primitive of function defined by \\
\begin{equation}
\displaystyle{\phi(u) = e^{\frac{u^{2}}{2}} \int_{u}^{\infty}e^{-\frac{s^{2}}{2}}ds}
\end{equation}
\n The function $\phi(u)$  has interesting properties, see {\color{blue}[23]}. \\

\n In particular\\

\n {\color{red}$\bullet_{1}$} For every non negative integer $n$, there exists a unique couple $(P_{n}, Q_{n})$ of polynomials that satisfy\\

\begin{equation}
\forall \,\, u \in \mathbb{R}, \quad \displaystyle{\phi^{(n)}(u) = P_{n}(u)\phi(u) - Q_{n}(u)}
\end{equation}
\n Moreover, these polynomials are defined, starting from $(P_{0}, Q_{0}) = (1, 0)$, by the recurrence relations
\begin{equation}
\displaystyle{\forall \,\, n \in \mathbb{N}, \quad P_{n+1} = xP_{n} + P_{n}^{'}}
\end{equation}
\begin{equation}
\displaystyle{\forall \,\, n \in \mathbb{N}, \quad Q_{n+1} = P_{n} + Q_{n}^{'}}
\end{equation}
\n and by  lemma 1 and proposition 3 of Kouba in {\color{blue}[23]} the sequence $\displaystyle{(P_{n}(x))_{n \in \mathbb{N}}}$ satisfies:\\
\begin{equation}
\displaystyle{\forall \, n \in \mathbb{N}, P_{n}(x) = e^{-\frac{x^{2}}{2}} \frac{d^{n}}{dx^{n}}e^{\frac{x^{2}}{2}}}
\end{equation} 

\n The polynomials $\displaystyle{P_{n}(x)}$ are linked to  the well known physicist's  Hermite polynomials by the following relation \\
\begin{equation}
\displaystyle{P_{n}(x) = (-\frac{i}{\sqrt{2}}) H_{n}(\frac{i}{\sqrt{2}} x)}
\end{equation}

\n {\color{red} $\rhd_{2}$}  In {\color{blue}[20]}, some similar properties of the generating function $\displaystyle{\varphi(u) = e^{u^{2}} \int_{u}^{\infty}e^{- s^{2}}ds}$ are given to study the ``Plasma dispersion function''.\\

\n In particular \\

\n {\color{red}$\bullet_{2}$ } The $n-th$ derivative of $\displaystyle{\varphi(u) = e^{u^{2}} \int_{u}^{\infty}e^{- s^{2}}ds}$ is given by:\\
\begin{equation}
\displaystyle{\varphi^{(n)}(u) = P_{n}(u) \varphi(u) - Q_{n}(u)}
\end{equation}

\n where $P_{n}$ and $Q_{n}$ are given by the following recurrence relations:\\
\begin{equation}
\left \{ \begin{array} {c} \displaystyle{ P_{0}(u) = 1, \quad P_{1}(u) = 2x} \quad \quad \quad \quad\\
\quad\\
\displaystyle{ Q_{0}(u) = 0, \quad Q_{1}(u) = 1}  \quad \quad \quad \quad\\
\quad\\
\displaystyle{P_{n+1}(u) = 2xP_{n}(u) + 2nP_{n-1}(u)}\\
\quad\\
\displaystyle{Q_{n+1}(u) = 2xQ_{n}(u) + 2nQ_{n-1}(u)}\\
\end{array} \right.
\end{equation}
\n Furthermore we have the following identity\\
\begin{equation}
\displaystyle{Q_{n+1}(u) P_{n}(u) - P_{n+1}(u) P_{n}(u) = (-2)^{n}n!}
\end{equation}
and the following expression for $P_{n}(u)$:\\
\begin{equation}
\displaystyle{P_{n}(u) = e^{-u^{2}}\frac{d^{n}}{du^{n}} e^{u^{2}}}
\end{equation}
\n In following lemma, we give an explicit expression of $v_{1}(y)$ and we show that it belongs to Bargmann space.\\
\end{remark}

\begin{lemma} ({\color{red} An explicit expression of $v_{1}(y)$})\\

\begin{equation}
\displaystyle{v_{1}(y) =  \sum_{n=0}^{\infty}(-1)^{n}a_{n}u_{2n + 1} - \sum_{n = 0}^{\infty} b_{n}u_{2(n +1)}(y)}
\end{equation}

where\\

$\displaystyle{a_{n} = \frac{\sqrt{(2n)! \frac{\pi}{2}}}{2^{n} n! \sqrt{2n + 1}}} $ and $\displaystyle{b_{n} = \frac{2^{n}n!}{\sqrt{2(n+1)(2n +1)!}}} $
\end{lemma}
\n {\color{red}{\bf Proof}}\\

\n Begining by written \\
 
\n $\displaystyle{v_{1} (y):= K_{0,\lambda} u_{1}(y) =  K_{0,\lambda}y =\frac{1}{\lambda} \int_{0}^{y}e^{\frac{u^{2}}{2}}du \int_{u}^{\infty}e^{-\frac{s^{2}}{2}}ds\}}$\\

\n $\displaystyle{ = \frac{1}{\lambda} \{\int_{0}^{y}e^{\frac{u^{2}}{2}}du [\int_{0}^{\infty}e^{-\frac{s^{2}}{2}}ds - \int_{0}^{u}e^{-\frac{s^{2}}{2}}ds]\}}$

\n $\displaystyle{=  -\frac{1}{\lambda}\int_{0}^{y}du \int_{0}^{u} e^{-\frac{1}{2}(s^{2} - u^{2})}ds }$  $\displaystyle{+ \frac{1}{\lambda} \int_{0}^{y}e^{\frac{u^{2}}{2}}du \int_{0}^{\infty}e^{-\frac{s^{2}}{2}}ds}$\\
\n $\displaystyle{= f(y) + g(y)}$\\

\n where \\

\n $\displaystyle{f(y) = -\frac{1}{\lambda}\int_{0}^{y}du \int_{0}^{u} e^{-\frac{1}{2}(s^{2} - u^{2})}ds }$  and  $\displaystyle{g(y) =  \frac{1}{\lambda} \int_{0}^{y}e^{\frac{u^{2}}{2}}du \int_{0}^{\infty}e^{-\frac{s^{2}}{2}}ds}$\\

\n  {\color{red}$\rhd_{1}$} Let  $\displaystyle{f(y) = -- \frac{1}{\lambda}\int_{0}^{y}du \int_{0}^{u} e^{-\frac{1}{2}(s^{2} - u^{2})}ds }$. \\

\n  Setting $\displaystyle{t = \frac{s}{u}}$ then we have\\

\n $\displaystyle{ \int_{0}^{u} e^{-\frac{1}{2}(s^{2} - u^{2})}ds = \int_{0}^{1}e^{-\frac{1}{2}(t^{2} - 1)u^{2}}udt }$\\

\n $\displaystyle{ =  \int_{0}^{1}\sum_{n=0}^{\infty}\frac{(-\frac{1}{2})^{n}(t^{2} - 1)^{n}u^{2n + 1}}{n!}}$\\

\n $\displaystyle{ =\sum_{n=0}^{\infty}\frac{u^{2n +1}}{2^{n} n!} \int_{0}^{1}(1 - t^{2})^{n}dt}$\\

\n $\displaystyle{ =\sum_{n=0}^{\infty}\frac{u^{2n +1}}{2^{n} n!} W_{n+1} }$ (where $\displaystyle{W_{n+1} = \frac{2^{2n}(n!)^{2}}{(2n + 1)!} \, \, \text{is Wallis formula})}$ \\

\n It follows that \\

\n $\displaystyle{ \int_{0}^{u} e^{-\frac{1}{2}(s^{2} - u^{2})}ds = \sum_{n=0}^{\infty}\frac{u^{2n +1}}{2^{n} n!}\frac{2^{2n}(n!)^{2}}{(2n + 1)!}}$ \\

\n $\displaystyle{ = \sum_{n=0}^{\infty}\frac{2^{n} n! \sqrt{(2n + 1)!}u_{2n +1}(u) }{(2n + 1)!}}$ \\

\n $\displaystyle{ = \sum_{n=0}^{\infty}\frac{2^{n} n!}{\sqrt{(2n + 1)!}} u_{2n+1}(u)}$ \\

\n Now as   $\displaystyle{ \int_{0}^{y}u_{2n+1}(u) du = \frac{1}{\sqrt{2(n+1)}} u_{2(n+1)}(y)}$ then we deduce that\\

\n $\displaystyle{f(y) = -\frac{1}{\lambda}\int_{0}^{y}du \int_{0}^{u} e^{-\frac{1}{2}(s^{2} - u^{2})}ds =  -\frac{1}{\lambda} \sum_{n=0}^{\infty}\frac{2^{n} n!}{\sqrt{2(n +1)(2n + 1)!}} u_{2(n+1)}(y)}$ \\

\n i.e.\\

\begin{equation}
\displaystyle{f(y) = -\frac{1}{\lambda}\sum_{n=0}^{\infty}a_{n} u_{2(n+1)}} \,\,\text{where}\,\, \displaystyle{a_{n} = \frac{2^{n} n!}{\sqrt{2(n +1)(2n + 1)!}}}
\end{equation}

\n  {\color{red}$\rhd_{2}$}  Now, we consider $\displaystyle{g(y) =  \frac{1}{\lambda} \int_{0}^{y}e^{\frac{u^{2}}{2}}du \int_{0}^{\infty}e^{-\frac{s^{2}}{2}}ds}$,  as  $\displaystyle{  \int_{0}^{\infty}e^{-\frac{s^{2}}{2}}ds = \sqrt{2\pi}}$, then we deduce that \\

\n $\displaystyle{g(y) = \frac{\sqrt{2\pi}}{\lambda} \int_{0}^{y}e^{\frac{u^{2}}{2}}du  = \frac{\sqrt{2\pi}}{\lambda} \int_{0}^{y}\sum_{n=0}^{\infty} \frac{u^{2n}}{2^{n}n!}}$$\displaystyle{= \frac{\sqrt{2\pi}}{\lambda} \sum_{n=0}^{\infty} \frac{y^{2n+1}}{2^{n}(2n + 1)n!}}$\\

\n As $\displaystyle{y^{n} = \sqrt{n!}u_{n}(y)}$, then it follows that\\

\n $\displaystyle{g(y)= \frac{\sqrt{2\pi}}{\lambda} \sum_{n=0}^{\infty} \frac{\sqrt{(2n + 1)!} u_{2n + 1}(y)}{2^{n}(2n + 1)n!}}$ \\

$\displaystyle{= \frac{\sqrt{2\pi}}{\lambda} \sum_{n=0}^{\infty} \frac{\sqrt{(2n)!} u_{2n + 1}(y)}{2^{n}n! \sqrt{2n+1}}}$\\

\n i.e.\\

\begin{equation}
\displaystyle{g(y) = \frac{\sqrt{2\pi}}{\lambda} \sum_{n=0}^{\infty} b_{n} u_{2n + 1}(y) \quad  \text{where}\,\,  b_{n} = \frac{\sqrt{(2n)!}}{2^{n}n! \sqrt{2n+1}}}
\end{equation}

\begin{corollary}
\quad\\
\n The function $v_{1}(y)$ belongs to Bargmann space, i.e. $\displaystyle{\sum_{n=0}^{\infty}\vert a_{n} \vert^{2} < \infty}$ and  $\displaystyle{\sum_{n=0}^{\infty}\vert b_{n} \vert^{2} < \infty}$\\
\end{corollary}

\n {\color{red}{\bf Proof}}\\
\begin{remark}
\quad\\
\n Let $\displaystyle{a_{n}^{2} = \frac{2^{2n}(n!)^{2}}{2(n+1)(2n + 1)!}}$, $\displaystyle{b_{n}^{2} = \frac{(2n)!}{2^{n}(n!)^{2}(2n + 1)}}$.\\

\n  Then by observing that $\displaystyle{(2n + 1)! = (2n)!(2n + 1)}$ and by setting $\displaystyle{c_{n} = \frac{(2n)!}{2^{n}(n!)^{2}}}$, we deduce that \\

\n $\displaystyle{b_{n}^{2} = \frac{c_{n}}{2n+1}}$, $\displaystyle{a_{n}^{2} = \frac{1}{2(n+1)(2n +1)c_{n}}}$ and\\

\begin{equation}
\displaystyle{a_{n}^{2}b_{n}^{2} = \frac{1}{2(n+1)(2n + 1)^{2}}\sim \frac{1}{n^{3}}}
\end{equation}

\n In particular\\

\begin{equation}
\displaystyle{a_{n}b_{n} \sim \frac{1}{n^{\frac{3}{2}}}}
\end{equation}
\n In bellow we show that $\displaystyle{b_{n} \sim \frac{1}{n^{\frac{3}{2}}}}$ to deduce the convergence of  $\displaystyle{\sum_{n=0}^{\infty}\vert b_{n} \vert^{2}}$ and of $\displaystyle{\sum_{n=0}^{\infty}\vert a_{n} \vert^{2} }$.\\
\end{remark}

\n The convergence of the series  $\displaystyle{ \sum_{n=1}^{\infty} \vert b_{n}\vert^{2}}$ requires Stirling's approximation, \\

\n $\displaystyle{ \lim\limits_{n \longrightarrow +\infty} \frac{1}{n!}(\frac{n}{e})^{n}\sqrt{2\pi n} = 1}$ and $\displaystyle{ \lim\limits_{n \longrightarrow +\infty} \frac{1}{(n!)^{2}}(\frac{n}{e})^{2n}2\pi n = 1^{2}}$ \\

\n We give two  methods to prove this convergence,  the first method by using  Stirling's approximation by establishing the convergence with limits and for the  second method, we use a upper bound of $n!$ by trapezoidal method and a lower bound of $n!$ by median point method.\\

\n {\color{red}(i) First method :} Establishing convergence with limits \\

\n Let $\displaystyle{a_{n}^{2} =   \frac{(2n)!}{2^{2n} (n!)^{2} (2n + 1)}}$ and  $\displaystyle{c_{n}^{2} = \frac{1}{n^{\frac{3}{2}}}}$. \\

\n Then\\

\n $\displaystyle{ \lim\limits_{n \longrightarrow +\infty} \frac{a_{n}^{2}}{c_{n}^{2}}}$  $\displaystyle{ = \lim\limits_{n \longrightarrow +\infty} =   \frac{ \frac{(2n)!}{2^{2n} (n!)^{2} (2n + 1)}}{\frac{1}{n^{3/2}}}} $\\

\n $\displaystyle{= \lim\limits_{n \longrightarrow +\infty} \frac{(2n)!n^{3/2}}{2^{2n} (n!)^{2} n} }$   by using $2n +1 \sim 2n$)\\

\n $\displaystyle{ = \lim\limits_{n \longrightarrow +\infty}  \frac{(2n)! \sqrt{n}}{2^{2n} (n!)^{2}} }$\\

\n $\displaystyle{ = \lim\limits_{n \longrightarrow +\infty}  \frac{(2n)! \sqrt{n}}{2^{2n} (n!)^{2}} {\color{red}\frac{1}{1^{2}}}}$\\

\n $\displaystyle{= \lim\limits_{n \longrightarrow +\infty}  \frac{(2n)! \sqrt{n}}{2^{2n} (n!)^{2}} {\color{red}\frac{ \frac{1}{(2n)!}(\frac{2n}{e})^{2n}\sqrt{4\pi n}}{( \frac{1}{n!}(\frac{n}{e})^{n}\sqrt{2\pi n})^{2}}}}$\\       

\n $\displaystyle{= \lim\limits_{n \longrightarrow +\infty} \frac{\sqrt{n}}{2^{2n}} {\color{red}\frac{(\frac{2n}{e})^{2n}\sqrt{4\pi n}}{(\frac{n}{e})^{2n}(2\pi n)}}}$\\  

\n $\displaystyle{= \lim\limits_{n \longrightarrow +\infty}  \frac{ \sqrt{n}}{2^{2n} } {\color{red}\frac{2^{2n}\sqrt{4\pi n}}{(2\pi n)}}}$\\ 

\n $\displaystyle{= \lim\limits_{n \longrightarrow +\infty} \frac{ \sqrt{n}{\color{red}\sqrt{4\pi n}}}{{\color{red}2\pi n}}}$\\ 

\n $\displaystyle{ = {\color{red}\frac{1}{\sqrt{\pi}}}}$\\

\n It follows that $\displaystyle{ a_{n}^{2} \sim \frac{1}{n^{\frac{3}{2}}}}$.\\

\n {\color{red}(ii) Second method :} lower bound and upper bound of $n!$\\

\n {\color{red}$\rhd_{1}$} By upper bound of $n!$ by using the trapezoidal method we get\\
\begin{equation}
\displaystyle{\quad \quad \quad n! \leq e n^{n} e^{-n}\sqrt{n} } \quad \quad \quad \quad  \quad  \quad  \quad  \quad {\color{blue} (\star_{1})}
\end{equation}
 This implies that \\
 \begin{equation}
\displaystyle{(2n)! \leq \alpha (2n)^{2n} e^{-2n}\sqrt{2n}}
\end{equation}

\n where $\alpha$ is a positive constant.\\

\n It follows that \\
 \begin{equation}
\displaystyle{\quad \quad \quad \quad \sqrt{(2n)!} \leq \sqrt{\alpha} (2n)^{n} e^{-n}(2n)^{\frac{1}{4}}}    \quad  \quad  \quad  \quad {\color{blue} (\star_{2})}
\end{equation}
\n {\color{red}$\rhd_{2}$} By lower bound of $n!$ by using the median point method we get\\
 \begin{equation}
 \displaystyle{\quad  \quad  \quad  \quad  \quad  \quad n! \geq c n^{n} e^{-n}\sqrt{n} \quad \quad  \quad  \quad  \quad  \quad  \quad  \quad} {\color{blue} (\star_{3})}
 \end{equation}
 
 \n where $c$ is a positive constant.\\
 
 \n This implies that \\
 \begin{equation}
 \displaystyle{ \quad  \quad  (2n)! \geq \beta (2n)^{2n} e^{-2n}\sqrt{2n} }
\end{equation}
 \n where $\beta$  is a positive constant\\
 
\n and\\
\begin{equation}
 \displaystyle{\quad \quad \sqrt{(2n)!} \geq \sqrt{\beta } (2n)^{n} e^{-n}(2n)^{\frac{1}{4}} \quad \quad \quad \quad \quad  \quad \quad \quad \quad  \quad {\color{blue} (\star_{4})}}
\end{equation}

\n From {\color{blue} ($\star_{1}$)} and {\color{blue} ($\star_{3}$)} we deduce that:\\
\begin{equation}
\displaystyle{ \quad \quad  c n^{n} e^{-n}\sqrt{n} \leq n! \leq e n^{n} e^{-n}\sqrt{n} } \quad \quad \quad \quad \quad \quad \quad \quad \quad \quad {\color{blue} (\star_{5})}
\end{equation}

\n From {\color{blue} ($\star_{2}$)} and {\color{blue} ($\star_{4}$)} we deduce that:\\
\begin{equation}
 \displaystyle{\sqrt{\beta } (2n)^{n} e^{-n}(2n)^{\frac{1}{4}}  \leq \sqrt{(2n)!}  \leq \sqrt{\alpha} (2n)^{n} e^{-n}(2n)^{\frac{1}{4}}}    \quad  \quad {\color{blue} (\star_{6})}
\end{equation}
\n Now from {\color{blue} ($\star_{5}$)} and {\color{blue} ($\star_{6}$)} we deduce that:\\
\begin{equation}
 \displaystyle{\frac{\sqrt{\beta } (2n)^{n} e^{-n}(2n)^{\frac{1}{4}}}{ e n^{n} e^{-n}\sqrt{n} }  \leq \frac{\sqrt{(2n)!} }{n!} \leq \frac{\sqrt{\alpha} (2n)^{n} e^{-n}(2n)^{\frac{1}{4}}}{c n^{n} e^{-n}\sqrt{n} }}    \quad  \quad {\color{blue} (\star_{7})}
\end{equation}
\n (Note in above denominators that $\displaystyle{e n^{n} e^{-n}\sqrt{n}}$ is  upper bound of $n!$  and $\displaystyle{c n^{n} e^{-n}\sqrt{n}}$ is lower bound of $n!$) \\

\n It follows that\\
\begin{equation}
 \displaystyle{\quad \quad \frac{\sqrt{\beta } 2^{n+\frac{1}{4}}}{e n^{\frac{1}{4}}}  \leq \frac{\sqrt{(2n)!} }{n!} \leq \frac{\sqrt{\alpha} 2^{n+\frac{1}{4}}}{c n^{\frac{1}{4}}}}   \quad \quad \quad  \quad \quad {\color{blue} (\star_{8})}
\end{equation}
\n Now as  $\displaystyle{b_{n} =  \frac{\sqrt{(2n)!}}{2^{n}n! \sqrt{2n+1}}}$, then we deduce that:\\
\begin{equation}
{\color{red} (i)}  \quad  \quad  \quad \quad  \quad  \displaystyle{\frac{\sqrt{\beta } 2^{\frac{1}{4}}}{e n^{\frac{1}{4}}\sqrt{2n+1}}  \leq b_{n} \leq \frac{\sqrt{\alpha} 2^{\frac{1}{4}}}{c n^{\frac{1}{4}}\sqrt{2n+1}}}    \quad  \quad  \quad  \quad   {\color{blue} (\star_{9})}
\end{equation}

\begin{equation}
{\color{red} (ii)}  \quad  \quad   \quad  \quad  \displaystyle{\sum_{n=1}^{\infty}b_{n}^{2} \leq  \frac{\alpha \sqrt{2}}{c^{2}} \sum_{n=1}^{\infty}\frac{1}{\sqrt{n} (2n + 1)} < \infty}  \quad \quad \quad {\color{blue} (\star_{10})}
\end{equation}

\n As  $\displaystyle{b_{n} \sim \frac{1}{n^{\frac{3}{4}}} \in \ell_{2}}$ and  $\displaystyle{a_{n}b_{n} \sim \frac{1}{n^{\frac{3}{2}}}}$ then  $\displaystyle{a_{n} \sim \frac{1}{n^{\frac{3}{4}}} \in \ell_{2}}$ and $v_{1} \in \mathbb{B}_{0}$.\\

\begin{corollary} 
\quad\\

$\displaystyle{v_{n}}$ belongs to Bargmann space for all $n=1, 2, ....$
\end{corollary}

\n {\bf {\color{red}Proof}}\\

\n As  $v_{n}$ satisfies the recurrence relation:\\
 $$\displaystyle{v_{n+1} = \frac{n-1}{\sqrt{ n(n+1)}}v_{n-1} + \frac{u_{n}}{\lambda n\sqrt{n+1}}} \quad \text{for all}\quad n = 1, 2, .... $$
\n where $u_{n}$ belongs to Bargmann space.\\

\n  Then we deduce that  $\displaystyle{v_{2} = \frac{u_{1}}{\lambda \sqrt{2}}}$, this implies that $v_{2} \in \mathbb{B}_{0}$ and by recurrence we deduce that $v_{n} \in \mathbb{B}_{0}$ for all $n = 1, 2, ...$\\

\begin{proposition} ({\color{red}Determination explicit of $v_{n}$ with respect $u_{n}$ and $v_{1}$})

\n Let \\

\n $\left \{ \begin{array}{c}\displaystyle{ v_{n+1} = A_{n}u_{n} + B_{n}v_{n-1}}\\  
\quad\\
\text{ where} \quad \quad \quad \quad \quad \quad \quad \quad \quad\\
 \displaystyle{A_{n} = \frac{1}{\lambda n\sqrt{n+1}}}\quad \quad \quad\\
 \text{and} \quad \quad \quad\quad \quad \quad \quad \quad \quad\\
 \displaystyle{B_{n} = \frac{n-1}{\sqrt{n(n+1)}}}\quad \quad \quad\\
\end{array} \right.$\hfill { } {\color{blue}($\star$)}
\quad\\

\n  Then we have\\

\begin{equation}
\displaystyle{v_{n} = P_{n-1} + \alpha_{n}v_{1}}
\end{equation}

\n with\\

$\left \{ \begin{array} {c}\displaystyle{\alpha_{n} = 0 \quad \quad \quad \quad \,\, \text{ if} \,\, n = 2p ; p \in \mathbb{N}} \quad \quad \quad \quad \quad\quad \quad \quad\quad \quad \quad\\
\quad\\
\displaystyle{\alpha_{2n +1} = \prod_{j=1}^{p}B_{2j} \quad \text{if} \,\, n = 2p + 1 ; p \in \mathbb{N}} \quad \quad \quad \quad \quad \quad\quad \quad\quad\\
\quad\\
P_{0} = 0  \quad \quad \quad \quad \quad \quad \quad  \quad \quad \quad \quad  \quad\quad \quad \quad\quad \quad \quad\quad \quad \quad\quad\\
\quad\\
\displaystyle{P_{1} = \frac{u_{1}}{\lambda\sqrt{2}}} \quad \quad \quad \quad \quad \quad \quad  \quad \quad \quad \quad \quad\quad \quad \quad\quad \quad \quad\quad \quad\\
\quad\:
\displaystyle{P_{n} =\frac{n-1}{\sqrt{n(n + 1)}} P_{n-2} +  \frac{u_{n}}{\lambda n\sqrt{n + 1}}} \quad \quad \quad \quad \quad \quad \quad \quad \quad \quad \quad \\
or \quad \quad \quad\quad \quad \quad\quad \quad \quad\quad \quad \quad \quad \quad \quad\quad \quad \quad\quad \quad \quad\quad \quad \quad\\ 
\displaystyle{P_{n} = B_{n} P_{n-2} + A_{n}u_{n}} \quad \quad \quad  \quad \quad \quad \quad \quad\quad \quad \quad\quad \quad \quad\quad \quad\\
\end{array} \right.$\\
\end{proposition}
\n {\color{red}{\bf Proof}}\\

\n $\displaystyle{ v_{2} = A_{1}u_{1} = P_{1} + \alpha_{2} v_{1};  P_{1} = A_{1}u_{1}}$ and $ \alpha_{2} = 0 $\\

\n $\displaystyle{ v_{3} = A_{2}u_{2} + {\color{red}B_{2}}v_{1} = P_{2} + \alpha_{3}v_{1}}$ ; $P_{2} = A_{2}u_{2}$ and $\alpha_{3} = B_{2}$ \\

\n $\displaystyle{ v_{4} = A_{3}u_{3} + B_{3}v_{2}}$ $\displaystyle{ = A_{3}u_{3} + B_{3}A_{1}u_{1} = P_{3} + \alpha_{4}v_{1}}$;  $P_{3} = A_{3}u_{3} + B_{3}A_{1}u_{1} = B_{3}P_{1} + A_{3}u_{3} $ and $\alpha_{4} = 0$ \\

\n $\displaystyle{ v_{5} = A_{4}u_{4} + B_{4}v_{3}}$ $\displaystyle{ = A_{4}u_{4} + B_{4}[ A_{2}u_{2} + {\color{red}B_{2}}v_{1}]}$ $\displaystyle{ = A_{4}u_{4} + B_{4}A_{2}u_{2} + {\color{red}B_{4}B_{2}}v_{1}}$ $\displaystyle{ = P_{4} + \alpha_{5}v_{1}}$: $\displaystyle{P_{4} =  A_{4}u_{4} + B_{4}A_{2}u_{2} = B_{4}P_{2} + A_{4}u_{4}}$ and $\displaystyle{\alpha_{5} = B_{4}B_{2} = \prod_{j=1}^{2} B_{2j}}$ \\

\n $\displaystyle{ v_{6} = A_{5}u_{5} + B_{5}v_{4}}$ $\displaystyle{ = A_{5}u_{5} + B_{5}[A_{3}u_{3} + B_{3}A_{1}u_{1}]}$ $\displaystyle{ = A_{5}u_{5} + B_{5}A_{3}u_{3} + B_{5}B_{3}A_{1}u_{1}}$ $\displaystyle{= P_{5}  + \alpha_{6}v_{1}}$; $\displaystyle{P_{5} = A_{5}u_{5} + B_{5}A_{3}u_{3} + B_{5}B_{3}A_{1}u_{1}}$  $\displaystyle{= B_{5}P_{3} + A_{5}u_{5}}$ and  $\displaystyle{\alpha_{6} = 0}$\\

\n $\displaystyle{ v_{7} = A_{6}u_{6} + B_{6}v_{5}}$ $\displaystyle{= A_{6}u_{6} + B_{6}[ A_{4}u_{4} + B_{4}A_{2}u_{2} + {\color{red}B_{4}B_{2}}v_{1}]}$ \\$\displaystyle{= A_{6}u_{6} + B_{6} A_{4}u_{4} + B_{6} B_{4}A_{2}u_{2} + {\color{red}B_{6}B_{4}B_{2}}v_{1}}$ $\displaystyle{  = P_{6} + \alpha_{7} v_{1}}$;  $\displaystyle{ P_{6} = B_{6}P_{4} + A_{6}u_{6}}$ and  $\displaystyle{\alpha_{7} = \prod_{j=1}^{3}B_{2j}}$\\

\n By recurrence with respect $p$, we suppose that\\

\n {\color{red}$\rhd_{2p}$} $\displaystyle{v_{2p} = P_{2p -1} + \alpha_{2p}v_{1}}$ such that $\displaystyle{P_{2p} = B_{2p} P_{2p -2} + A_{2p}u_{2p}}$ and $\displaystyle{\alpha_{2p} = 0}$.\\

\n {\color{red}$\rhd_{2p+1}$} $\displaystyle{v_{2p+1} = P_{2p } + \alpha_{2p+1}v_{1}}$ such that $\displaystyle{P_{2p+1} = B_{2p+1} P_{2p -1} + A_{2p+1}u_{2p+1}}$ and $\displaystyle{\alpha_{2p+1} = \prod_{j}^{p}B_{2j}}$.\\

\n Then we deduce that\\

\n {\color{red} $\bullet_{2(p + 1)}$} $\displaystyle{ v_{2(p+1)} = A_{2p + 1} u_{2p+1} + B_{2p+1}v_{2p}} $ \\

\n $\displaystyle{ = A_{2p + 1} u_{2p+1} + B_{2p+1}[P_{2p -1} + \alpha_{2p}v_{1}]}$\\

\n $\displaystyle{=  A_{2p + 1} u_{2p+1} + B_{2p+1}P_{2p -1} + \alpha_{2p}B_{2p+1}v_{1}}$ \\

\n  $\displaystyle{ = B_{2p+1}P_{2p -1} + A_{2p + 1} u_{2p+1}}$ (because $\alpha_{2p} = 0$) \\

\n $\displaystyle{= B_{2p+1}P_{2p -1} + A_{2p + 1} u_{2p+1} + \alpha_{2(p+1)}v_{1}}$. \\

\n This implies \\

\n $\displaystyle{\alpha_{2(p+1)} = 0}$\\

\n  and\\

\n  $\displaystyle{P_{2p+1} = B_{2p+1} P_{(2p + 1) -2} + A_{2p + 1} u_{2p+1} }$.\\

\n i.e.\\

\n  $\displaystyle{v_{2(p+ 1) } =  P_{2p+1} + \alpha_{2(p+1)}v_{1}}$\\

\n {\color{red} $\bullet_{2(p + 1) + 1}$} $\displaystyle{v_{2(p+1) + 1} = A_{2(p+1)} u_{2(p+1)} +  B_{2(p+1)}v_{2p +1}}$\\

\n $\displaystyle{= A_{2(p+1)} u_{2(p+1)} +  B_{2(p+1)}[ P_{2p } + \alpha_{2p+1}v_{1}]}$ \\
 
\n $\displaystyle{= A_{2(p+1)} u_{2(p+1)} +  B_{2(p+1)}P_{2p } + B_{2(p+1)}\alpha_{2p+1}v_{1}]}$\\

\n  $\displaystyle{= A_{2(p+1)} u_{2(p+1)} +  B_{2(p+1)}P_{2p } + B_{2(p+1)}\prod_{j=1}^{p}B_{2j}v_{1}}$\\

\n  $\displaystyle{=  B_{2(p+1)}P_{2p }  + A_{2(p+1)} u_{2(p+1)} + \prod_{j=1}^{p+1}B_{2j}v_{1}}$\\

\n Then we deduce that\\

\n $\displaystyle{\alpha_{2(p+1) +1} = \prod_{j=1}^{p+1}B_{2j}v_{1}}$\\

\n  and\\

\n  $\displaystyle{P_{2(p+1)} =  B_{2(p+1)}P_{2p }  + A_{2(p+1)} u_{2(p+1)} }$.\\

\n i.e.\\

\n  $\displaystyle{v_{2(p+ 1) + 1} =  P_{2(p+1) }   + \alpha_{2(p+1) +1}v_{1}}$\\

\begin{lemma}
\n Let $\displaystyle{u_{n}(y) = \frac{y^{n}}{\sqrt{n!}}}$  with norm $\vert\vert u_{n} \vert\vert = 1$ and $P_{n}(y)$  is a polynomial which satisfy :\\

$$\displaystyle{P_{n}(y) = B_{n} P_{n-2} (y) + A_{n}u_{n}(y)}$$

\n where  $\displaystyle{B_{n} = \frac{n-1}{\sqrt{n(n+1)}}}$ and $\displaystyle{A_{n} = \frac{1}{\lambda n\sqrt{n+1}}}$\\

\n Then \\

\n (i) $P_{n-2} $ is orthogonal to $u_{n}$.\\

\n (ii) Let $\displaystyle{p_{n} = \vert\vert P_{n} \vert\vert^{2}}$ then  $\displaystyle{p_{n} = B_{n}^{2} p_{n-2} + A_{n}^{2}}$\\

\n (iii) $\displaystyle{p_{n} = O(\frac{1}{n^{\frac{3}{2}}})}$.\\
\end{lemma}

\n {\color{red}{\bf Proof}}\\

\n (i) As the degree of $P_{n-2}$ is $n - 2$ then $\displaystyle{ < P_{n-2} , u_{n} > = 0}$\\

\n (ii) We have \\

\n $\displaystyle{p_{n} = \vert\vert P_{n} \vert\vert^{2} = < B_{n} P_{n-2} (y) + A_{n}u_{n}(y)\,, \, B_{n} P_{n-2} (y) + A_{n}u_{n}(y) >}$

\n $\displaystyle{= < B_{n} P_{n-2} (y) , B_{n} P_{n-2} (y) >  + < B_{n} P_{n-2} (y), A_{n}u_{n}(y) >}$\\

 \n $\displaystyle{ + < A_{n}u_{n}(y) , B_{n} P_{n-2} (y) > + < A_{n}u_{n}, A_{n}u_{n} > }$\\
 
\n By using (i) and $\vert\vert u_{n} \vert\vert = 1$, we  deduce that  $\displaystyle{p_{n} = B_{n}^{2} p_{n-2} + A_{n}^{2}}$\\

\n (iii) is deduced from (ii)\\

\begin{proposition} ({\color{red}Compactness of $K_{0,\lambda}$})
\quad\\
\n Let $\displaystyle{u = \sum_{n=1}^{\infty}c_{n}u_{n}}$, $\displaystyle{K_{0,\lambda}u = \sum_{n=1}^{\infty}c_{n} v_{n} = \sum_{n=1}^{\infty}c_{n}P_{n-1} + c_{1}v_{1}}$ where $c_{1}$ is a constant and $\displaystyle{K_{0,\lambda, m}u = \sum_{n=1}^{m}c_{n}P_{n-1} + c_{1}v_{1}}$.\\

\n  Then \\

\n (i) $\displaystyle{K_{0,\lambda, m} \longrightarrow K_{0,\lambda} }$ as $m \longrightarrow \infty$ (in operator norm).\\

\n (ii) $K_{0, \lambda}$ is compact.\\
\end{proposition}

\n {\color{red}{\bf Proof}}\\

\n (i) Let $\displaystyle{u = \sum_{n=1}^{\infty}c_{n}u_{n}}$ then $\displaystyle{K_{0,\lambda}u = \sum_{n=1}^{\infty}c_{n}K_{0,\lambda} u_{n} = \sum_{n=1}^{\infty}c_{n} v_{n} =  \sum_{n=1}^{\infty}c_{n}(P_{n-1} + \alpha_{n}v_{1})}$\\

\n Let $\displaystyle{u = \sum_{n=1}^{\infty}c_{n}P_{n-1} + cv_{1}}$ where $c $ is a constant.\\

\n Now, let  $\displaystyle{p_{n} = \vert\vert P_{n} \vert\vert^{2}}$ then it follows by using Cauchy-Schwarz inequality that:\\

\n $\displaystyle{\vert\vert \sum_{n= m+1}^{\infty}c_{n}P_{n-1}\vert\vert \leq \sum_{n= m+1}^{\infty}\vert c_{n}\vert.\vert\vert P_{n-1}\vert\vert}$\\

\n $\displaystyle{\leq  (\sum_{n= m+1}^{\infty}\vert c_{n}\vert^{2})^{\frac{1}{2}}. (\sum_{n= m+1}^{\infty}\vert\vert P_{n-1}\vert\vert^{2})^{\frac{1}{2}}}$\\

\n $\displaystyle{\leq  (\sum_{n= m+1}^{\infty}\vert < u, u_{n} > \vert^{2})^{\frac{1}{2}}. (\sum_{n= m}^{\infty}\vert\vert p_{n}\vert\vert)^{\frac{1}{2}}}$\\

\n $\displaystyle{\leq \vert\vert u \vert\vert. (\sum_{n = m}^{\infty}\vert\vert p_{n}\vert\vert)^{\frac{1}{2}}}$\\

\n Then $\displaystyle{K_{0,\lambda, m} \longrightarrow K_{0,\lambda} }$ as $m \longrightarrow \infty$ (in operator norm).\\

\n (ii) As $K_{0,\lambda}$ is an operator norm limit of finite rank operators then $K_{0,\lambda}$ is a compact operator. \\

\section{{\color{red} On the complete indeterminacy of generalized Heun's operator $\displaystyle{H^{p,m} = A^{*p}(A^{m} + A^{*m})A^{p}, p, m = 1,2, ..}$ in Bargmann space}}

\n We recall that  in Bargmann representation, the standard Bose annihilation and creation operators are defined by:\\

\begin{equation}
\displaystyle{A\varphi(z) = \varphi^{'}(z) \,\, \text{with domain} \, \, D(A) = \{\varphi \in \mathbb{B} ; A\varphi \in \mathbb{B}\}}
\end{equation}
\begin{equation}
\displaystyle{A^{*}\varphi(z) = z \varphi(z) \,\, \text{with domain} \, \, D(A^{*}) = \{\varphi \in \mathbb{B} ; A^{*}\varphi \in \mathbb{B}\}}
\end{equation}

\n It follows from (3.1) and (3.2) that the action of the operator $\displaystyle{A^{*^{r}}A^{s}}$ $(r \in \mathbb{N}, s \in \mathbb{N})$ on an element $\displaystyle{\varphi \in \mathbb{B}; \varphi(z) = \sum_{k=0}^{\infty}a_{k}e_{k}(z)}$ where $\displaystyle{ e_{k}(z) = \frac{z^{k}}{\sqrt{k!}}}$ is given by :\\

\begin{equation}
\displaystyle{A^{*^{r}}A^{s}\varphi(z) = \sum_{k=0}^{\infty}k(k-1) ....... (k -s +1)a_{k}\frac{z^{k+r -s}}{\sqrt{k!}}}
\end{equation}
\n or\\
\begin{equation}
\displaystyle{A^{*^{r}}A^{s}\varphi(z) = \sum_{k=0}^{\infty}(k+s - r)(k+s - r-1) .... (k-r +1)a_{k+s - r}\frac{z^{k}}{\sqrt{(k+s - r)!}}}
\end{equation}

\n If $s \geq r$, we have $\displaystyle{\sqrt{(k+s - r)!} = \sqrt{k!}\sqrt{k+1} ...... \sqrt{k + s  -r}}$. This implies that\\
\begin{equation}
\displaystyle{A^{*^{r}}A^{s}\varphi(z) = \sum_{k=0}^{\infty}\{\sqrt{(k+s - r)}\sqrt{(k+s - r-1} .... \sqrt{k+1} ...... (k-r +1)a_{k+s - r}\} e_{k}(z)}
\end{equation} 
\n Let $\displaystyle{u_{k} = \sqrt{(k+s - r)}\sqrt{(k+s - r-1} .... \sqrt{k+1} .... (k-r +1)a_{k+s - r}}$ then we give the bellow {\color{red}obvious lemma} that we will use in the following of this section.\\

\begin{lemma} ({\color{blue}[17]}, p. 1487)\\
\n (i) If $s \geq r$ then $\displaystyle{u_{k} \sim k^{\frac{s+r}{2}}}$\\

\n (ii)  If $s \leq r$ then $\displaystyle{u_{k} \sim k^{\frac{s+r}{2}}}$\\

\n (iii) If $r = p$ and s = $p + m$ then $\displaystyle{u_{k} \sim k^{\frac{2p+m}{2}} = k^{p + \frac{m}{2}} }$\\
\end{lemma}
\n {\color{red}{\bf Proof}}\\

\n (i) We observe that $\displaystyle{u_{k} \sim k^{\frac{s-r}{2}}k^{r}}$ then it follows that $\displaystyle{u_{k} \sim k^{\frac{s+r}{2}}}$\\

\n (ii) In a similar manner, we obtain $\displaystyle{u_{k} \sim k^{\frac{s+r}{2}}}$ when $s \leq r$\\

 \n (iii) is obvious particular case of (i). The proof of this Lemma  is complete.\\
 
\n  Now the action of the operator $H^{p,m}$ on an element $\varphi \in \mathbb{B}$ is given by:\\
 
\n $\displaystyle{H^{p,m}\varphi(z) = A^{*^{p}}A^{m + p}\varphi(z) + A^{*^{p + m}}A^{p}\varphi(z)}$\\
\n or\\
\begin{equation}
\displaystyle{H^{p,m}\varphi(z) = \sum_{k=0}^{\infty}k(k-1)...(k-p - m +1)a_{k}\frac{z^{k-m}}{\sqrt{k!}}+\sum_{k=0}^{\infty}k(k-1)...(k - p + 1)a_{k}\frac{z^{k+m}}{\sqrt{k!}}}
\end{equation}
\n We observe that $H^{p,m}$ is polynomial in $(A^{*}, A)$ of degree $2p + m$ and the differential operations $A^{*}$ and $A$ act on the functions $e_{k}(z)$ according to the formulas\\
\begin{equation}
\displaystyle{A^{*}e_{k} = \sqrt{k+1}e_{k+1},\,\, \text{and} \,\, Ae_{k} = \sqrt{k}e_{k-1}; e_{-1} = 0, k = 0, 1 , . . .}
\end{equation}

\n It follows from (3.7) that\\
\begin{equation}
\left\{\begin{array}{c}\displaystyle{H^{p,m}e_{k} = 0}; \quad k < p \quad\quad\quad\quad\quad\quad\quad\quad\quad\quad\quad\quad\quad\quad\quad\quad\quad\\
\quad\\
\displaystyle{H^{p,m}e_{k} = \frac{\sqrt{k!(k + m)!}}{(k - p)!}e_{k+m}}; \quad p\leq k < p+m \quad\quad\quad\quad\quad\quad\\
\quad\\
\displaystyle{H^{p,m}e_{k}=\frac{\sqrt{k!(k - m)!}}{(k - p - m)!}e_{k-m}+\frac{\sqrt{k!(k + m)!}}{(k - p)!}e_{k + m}};k \geq p+m \\
\end{array} \right.  \\
\end{equation}

\n Thus from (3.8), if we denote \\
\begin{equation}
\displaystyle{\mathbb{B}_{p} = \{\phi \in \mathbb{B}; \phi(0) = \phi^{'}(0) = ..... = \phi^{(p-1)}(0) = 0 \}}
\end{equation}
\n then this space is generated by $\displaystyle{\{e_{p}, e_{p+1}, .... \}}$ and the matrix representation of the minimal operator $\mathbb{H}$ generated by the expression $H^{p,m}$ in the basis $\displaystyle{e_{k}; k = p, p+1, ... }$ is given by the symmetric Jacobi matrix $\mathbb{H}$ witch has only two nonzero diagonals. Namely,  its numerical entries are the matrices $\displaystyle{H_{ij}}$ of order $m$ defined by:\\

\begin{equation}
\mathbb{J} = \left\{\begin{array}{c}\displaystyle{H_{i,i} =H_{i,j} = O \quad; if \quad \mid i-j\mid > 1}\quad (i,j = 1, 2, ...) \quad \quad\quad\quad\quad \quad\quad\quad\quad\\
 \text{where}\,\, O \,\, \text{is the zero} \,\, m\times m\,  \text{matrix}   \quad\quad\quad\quad\quad\quad\quad\quad \quad\quad\quad\quad\quad \quad\quad\quad\quad\\ 
 \quad\\
  \displaystyle{H_{i+1,i} =H_{i, i+1}}  \quad\quad\quad\quad \quad\quad\quad\quad \quad\quad\quad\quad \quad \quad\quad\quad\quad \quad\quad\quad\quad  \quad\quad\quad\quad\\
  \text{where}\, H_{i, i+1}\,  \text{is diagonal} \,  m\times m\,  \text{ matrix such that its numerical entries  are }\\
  \quad\\
\displaystyle{\beta_{k}^{i} = \frac{\sqrt{k!(k + m)!}}{(k - p)!};\quad (i-1)m + 1\leq k \leq im}  \quad\quad\quad\quad\quad\quad\quad\quad \quad\quad\quad\quad \quad\\ 
\end{array}\right.\\
\end{equation}

\n Let $\displaystyle{A_{i}}$ and $\displaystyle{B_{i} = B_{i}^{*}}$ $(i= 1, 2, .......)$ be $m\times m$ matrices whose entries are complex numbers then the matrix (3.10) is a particular case of the infinite matrix whose general form\\

\begin{equation}
\mathfrak{J}= \left(
  \begin{array}{ c c c c c c c c c }
     A_{1} & B_{1} & O & .&.&.&.& \\
     B_{1}^{*}  & A_{2} & B_{2} &\ddots & .&.&.\\
    O &B_{2}^{*} & A_{3} & B_{3}& \ddots &.&.\\
    . & \ddots & \ddots & \ddots & \ddots & \ddots & .&\\
     . & . & \ddots & \ddots & \ddots  & \ddots& O & \\
     . & . & . & \ddots & \ddots  & \ddots & \ddots &\\
     .   &. & . &.& O & \ddots & \ddots& \\
  \end{array} \right).
  \end{equation}

 \n  where $O$ is the zero $m\times m$ matrix and the asterisk denotes the adjoint matrix.\\

 \n  Let $l_{m}^{2}(\mathbb{N})$ be the Hilbert space of infinite sequences $\phi = (\phi_{1}, \phi_{2},.....,\phi_{i}, ......)$ with the inner product $\displaystyle{< \phi, \psi > = \sum_{i=1}^{\infty}\phi_{i}\overline{\psi}_{i}}$\\

\n   where $\displaystyle{\phi_{i} = (\phi_{i}^{1}, \phi_{i}^{2},.....,\phi_{i}^{m}) \in \mathbb{C}^{m}}$ and $\displaystyle{\phi_{i}\overline{\psi}_{i} = \sum_{j=1}^{m}\phi_{i}^{j}\overline{\psi}_{i}^{j}}$\\

\n  The matrix $\mathfrak{J}$ defines a symmetric operator $\mathfrak{T}$ in $l_{m}^{2}(\mathbb{N})$  according to the formula \\
\begin{equation}
  \displaystyle{(\mathfrak{T}\phi)_{i} = B_{i-1}\phi_{i-1} + A_{i}\phi_{i} + B_{i}\phi_{i+1}, i = 1, 2, .....}
\end{equation}
\n where $\displaystyle{\phi_{0} = (\phi_{0}^{1}, \phi_{0}^{2},.....,\phi_{0}^{m}) = (0, 0, .....,0)}$\\

\n then for our operator, we have:\\

\begin{equation}
\displaystyle{(H\phi)_{i} = H_{i,i-1}\phi_{i-1}  + H_{i,i+1}\phi_{i+1}, i = 1, 2, .....}
\end{equation}

\n where $\displaystyle{\phi_{0} = (\phi_{0}^{1}, \phi_{0}^{2},.....,\phi_{0}^{m}) = (0, 0, .....,0)}$\\

\begin{remark} ({\color{red}Recall of some classical results})

\n (1) The closure $\mathbb{T}$ with domain $D(\mathbb{T})$ of the operator $\mathfrak{T}$ is the minimal  closed symmetric operator generated by the expression (3.12) and the boundary condition $\phi_{0} = 0$\\

\n (2) According to Berezanskii, by Chap VII, $\S \, 2$ {\color{blue}[4]} , it is well known that the deficiency numbers $n_{+}$ and  $n_{-}$\, of the operator $\mathbb{T}$ satisfy the inequalities $0 \leq  n_{+} \leq m$ and $0 \leq  n_{-} \leq m$\\

\n where  $n_{+}$ is the dimension of $\mathfrak{M}_{z} = (\mathbb{T} - zI)D(\mathbb{T}); \mathfrak{I}m z \neq 0$ and $n_{-}$  is the dimension of the eigensubspace $\mathfrak{N}_{\overline{z}}$ corresponding to the eigenvalue $\overline{z}$ of the operator $\mathbb{T}$\\

\n (3) According to Krein {\color{blue}[24]} that the operator $\mathbb{T}$ is said completely indeterminate if $n_{+} = n_{-} = m$ and to Kostyuchenko-Mirsoev {\color{blue}[22]} that the completely indeterminate case holds for the operator $\mathbb{T}$ if and only if all solutions of the vector equation\\
\begin{equation}
(\mathfrak{T}\phi)_{i} = z\phi_{i} \quad i=1, 2, .... .
\end{equation}

\n for $z  = 0$ belongs to $l_{m}^{2}(\mathbb{N})$\\

\n (4)  In {\color{blue}[22]}, Kostyuchenko and Mirzoev gave some tests for the complete indeterminacy of a Jacobi matrix $\mathfrak{J}$ in terms of entries $A_{i}$ and $B_{i}$ of that matrix. \\

\n (5) In the following, we give two lemmas witch permit us to show the complete indeterminacy of generalized Heun operator  $\displaystyle{H^{p,m} = a^{*^{p}} (a^{m} + a^{*^{m}})a^{p}}$ in Bargmann space.\\
\end{remark}

\n For $m = 1, 2, ...$, let $\mathbb{C}^{m}$ be the euclidean $m-$dimentional space and
$\displaystyle{B_{i} = H_{i,i+1}}$ be the diagonal $ m\times m$ matrix such that its numerical entries are \\

\n $\displaystyle{\beta_{k}^{i} = \frac{\sqrt{k!(k + m)!}}{(k - p)!};\quad (i-1)m + 1\leq k \leq im} (i=1,2, .....)$.\\

\n By $\mid\mid . \mid\mid$ we denote the spectral matrix norm, then we have:\\

\n {\color{red}$\rhd_{1}$} $\displaystyle{\mid\mid B_{i} \mid\mid = \frac{\sqrt{(im)![(i+1)m]!}}{(im - p)!} \sim i^{p+\frac{m}{2}}}$ \\

\n {\color{red}$\rhd_{2}$}  $\displaystyle{\mid\mid B_{i}^{-1} \mid\mid = \frac{1}{\mid\mid B_{i} \mid\mid}}$\\

\begin{lemma}

\n Let $\displaystyle{B_{i} = H_{i,i+1}}, (i = 1, 2, ...)$ be the diagonal $ m\times m $ matrix ($m = 1, 2, ...$) such that its numerical entries are \\

$$\displaystyle{\beta_{k}^{i} = \frac{\sqrt{k!(k + m)!}}{(k - p)!};\quad (i-1)m + 1\leq k \leq im} (i=1,2, .....)$$

\n then the following inequality holds\\
\begin{equation}
\displaystyle{\mid\mid B_{i-1}\mid\mid \mid\mid B_{i+1}\mid\mid \leq \frac{1}{\mid\mid B_{i}^{-1}\mid\mid^{2}}}
\end{equation}
holds starting from some $i \geq m$\\
\end{lemma}

\n {\color{red}{\bf Proof}}\\

\n By using the lemma 3.1 or the behavior of Gamma function $\Gamma(x)$ as \\$\mathcal{R}e x \rightarrow +\infty$ given by Stirling's formula $\displaystyle{\Gamma(x)\sim\sqrt{2\pi}e^{-x}x^{x-\frac{1}{2}}}$, we deduce that \\

 $$\displaystyle{\beta_{k}^{i} \sim k^{p+\frac{m}{2}}}\,\, \text{as}\,\, k \rightarrow +\infty$$

\n As $\displaystyle{\mid\mid B_{i}\mid\mid = \beta_{im}^{i} = \frac{\sqrt{(im)!(m(i+1))!}}{(im - p)!}\sim (im)^{p+\frac{m}{2}} = (i)^{p+\frac{m}{2}}(m)^{p+\frac{m}{2}}}$

 \n then  $\displaystyle{\mid\mid B_{i-1}\mid\mid \sim(i-1)^{p+\frac{m}{2}}(m)^{p+\frac{m}{2}}}$, $\displaystyle{\mid\mid B_{i+1}\mid\mid \sim(i+1)^{p+\frac{m}{2}}(m)^{p+\frac{m}{2}}}$ and \\
\n $\displaystyle{\mid\mid B_{i-1}\mid\mid}$.$\displaystyle{\mid\mid B_{i+1}\mid\mid}$ $\displaystyle{\sim (i)^{2p+ m}(1- \frac{1}{i^{2}})^{p+\frac{m}{2}}(m)^{2p+m}}$\\

\n Now as  $\displaystyle{(1 - \frac{1}{i^{2}})^{p+\frac{m}{2}} \leq 1}$ then \\

\n  $\displaystyle{\mid\mid B_{i-1}\mid\mid}$ $\displaystyle{\mid\mid B_{i+1}\mid\mid} \leq $ $\displaystyle{\mid\mid B_{i}\mid\mid^{2}}$ and as $\displaystyle{\mid\mid B_{i}^{-1} \mid\mid = \frac{1}{\mid\mid B_{i} \mid\mid}}$ then  (3.15) holds\\

\begin{lemma}

\n Let $\displaystyle{B_{i} = H_{i,i+1}}, (i = 1, 2, ...)$ be the diagonal $ m\times m$ matrix , $(m = 1, 2, ...)$ such that its numerical entries are \\

$$\displaystyle{\beta_{k}^{i} = \frac{\sqrt{k!(k + m)!}}{(k - p)!};\quad (i-1)m + 1\leq k \leq im}$$

\n then if $2p + m > 2$ the following inequality holds\\
\begin{equation}
\displaystyle{\sum_{i=1}^{+\infty}\frac{1}{\mid\mid B_{i}\mid\mid} < +\infty}
\end{equation}
\end{lemma}

\n {\color{red}{\bf Proof}}\\

\n As $\displaystyle{\mid\mid B_{i}\mid\mid\sim i^{p + \frac{m}{2}}}$ then  if $p + \frac{m}{2} > 1$ , the serie $\displaystyle{\sum_{i=1}^{+\infty}\frac{1}{i^{p + \frac{m}{2}}}} < +\infty$, it follows that (3.16) holds.\\

\n Now, we prove the following theorem\\

\begin{theorem}

\n If $p + \frac{m}{2} > 1$ then the operator $\mathbb{H}$  is completely indeterminate and its deficient numbers satisfy the conditions $n_{+} =  n_{-} = m$.\\
\end{theorem}
\n {\color{red}{\bf Proof}}\\

\n By applying the results of Kostyuchenko and Mirsoev {\color{blue}[22]} to our operator then the completely indeterminate case holds for the operator $\mathbb{H}$ if and only if all solutions of the vector equation\\

\n $\displaystyle{ B_{i-1}\phi_{i-1}  + B_{i}\phi_{i+1} = \lambda \phi_{i}\quad (i = 1, 2, .....)}$\\

\n for $\lambda = 0$ belongs to $l_{m}^{2}(\mathbb{N})$\\

\n where $\displaystyle{B_{i} = H_{i,i+1}}$ is the diagonal $ m\times m$ matrix such that its numerical entries are given by
$\displaystyle{\beta_{k}^{i} = \frac{\sqrt{k!(k + m)!}}{(k - p)!};\quad (i-1)m + 1\leq k \leq im}$\\

\n Now from (3.13) we consider the system\\

\n $\displaystyle{ B_{i-1}\phi_{i-1}  + B_{i}\phi_{i+1} = 0 \quad (i = 1, 2, .....)}$\\

\n where $\displaystyle{\phi_{0} = (\phi_{0}^{1}, \phi_{0}^{2},.....,\phi_{0}^{m}) = (0, 0, .....,0)}$\\

\n As $\displaystyle{B_{i}^{-1},\quad (i = 1, 2, .....)}$ exist we deduce that the solutions of the above equation have the following explicit form\\

\n If $i= 2j, \quad (j = 1, 2, ....)$ we have\\

$\displaystyle{\phi_{2j} = 0 }$ and $\displaystyle{\phi_{2j+1} = -(1)^{j}B_{2j}^{-1}B_{2j-1}\times B_{2j-2}^{-1}B_{2j-3} ...... \times B_{2}^{-1}B_{1}\phi_{1}}$\\

\n This solution belongs to $l_{2}(\mathbb{N})$ if $\displaystyle{\sum_{j=1}^{+\infty}\mid\mid B_{2j}^{-1}B_{2j-1}\times B_{2j-2}^{-1}B_{2j-3} ...... \times B_{2}^{-1}B_{1}\mid\mid^{2} < + \infty}$\\

\n and\\

\n if $i= 2j-1, \quad (j = 1, 2, ....)$ we have\\

\n $\displaystyle{\phi_{2j-1} = 0 }$ and $\displaystyle{\phi_{2j} = -(1)^{j}B_{2j-1}^{-1}B_{2j-2}\times B_{2j-3}^{-1}B_{2j-4} ...... \times B_{3}^{-1}B_{2}\phi_{2}}$\\

\n This solution belongs to $l_{2}(\mathbb{N})$ if $\displaystyle{\sum_{j=1}^{+\infty}\mid\mid B_{2j-1}^{-1}B_{2j-2}\times B_{2j-3}^{-1}B_{2j-4} ...... \times B_{3}^{-1}B_{2}\mid\mid^{2} < + \infty}$\\

\n Then the solution generated by the above solutions belongs to $l_{2}(\mathbb{N})$ if \\
\begin{equation}
\displaystyle{\sum_{j=1}^{+\infty}\mid\mid B_{2j-1+\epsilon}^{-1}B_{2j-2+\epsilon}\times ...... \times B_{3+\epsilon}^{-1}B_{2+\epsilon}B_{1+\epsilon}^{-1}B_{\epsilon}\mid\mid^{2} < + \infty}
\end{equation}

\n where $\epsilon = 0$ or $\epsilon = 1$ and $\displaystyle{B_{0} = B_{1}^{-1}}$\\

\n Now as \\

\n $\displaystyle{\mid\mid B_{2j-1+\epsilon}^{-1}B_{2j-2+\epsilon}\times ...... \times B_{3+\epsilon}^{-1}B_{2+\epsilon}B_{1+\epsilon}^{-1}B_{\epsilon}\mid\mid^{2} \quad \leq}$\\

\n $\displaystyle{\mid\mid B_{2j-1+\epsilon}^{-1}\mid\mid^{2}\mid\mid B_{2j-2+\epsilon}\mid\mid^{2}\times ...... \times \mid\mid B_{3+\epsilon}^{-1}\mid\mid^{2}\mid\mid B_{2+\epsilon}\mid\mid^{2}\mid\mid B_{1+\epsilon}^{-1}\mid\mid^{2}\mid\mid B_{\epsilon}\mid\mid^{2}}$\\

\n then it follows from (3.15) of lemma 3.3 that\\

\n $\displaystyle{\mid\mid B_{2j-1+\epsilon}^{-1}\mid\mid^{2}\times ...... \times \mid\mid B_{3+\epsilon}^{-1}\mid\mid^{2}\mid\mid B_{1+\epsilon}^{-1}\mid\mid^{2} \quad \leq }$\\

\n $\displaystyle{\frac{1}{\mid\mid B_{2j-2+\epsilon}\mid\mid^{2}\times ...... \times \mid\mid B_{2+\epsilon}\mid\mid^{2}\mid\mid B_{1+\epsilon}^{-1}\mid\mid^{2}\mid\mid B_{\epsilon}\mid\mid \mid\mid B_{2j+\epsilon}\mid\mid }}$\\

\n and consequently the general term of the series (3.17) do not exceed \\ 

\n $\displaystyle{\frac{\mid\mid B_{\epsilon}\mid\mid }{\mid\mid B_{2j+\epsilon}\mid\mid}}$
and (3.16) of the lemma 3.4. ensures the convergence of the series $\displaystyle{\sum_{j=1}^{+\infty}\frac{1}{\mid\mid B_{2j+\epsilon}\mid\mid}}$. The  proof of Theorem 3.5 is complete.\\

\n We end this section by following remark:\\

\begin{remark}
\n (i) Let $\mathbb{U}$ be an operator defined by $\displaystyle{\mathbb{U}e_{k} = e_{k-m+1}}$ and $\mathbb{H}$ be an operator defined by $\displaystyle{\mathbb{H}e_{k} = A^{*^{p}}A^{p+m}\mathbb{U}}$ $(p, m = 1, 2, ....)$ then\\
\begin{equation}
\displaystyle{\mathbb{H}e_{k} = \omega_{k-1}^{m,p}e_{k-1} \,\, \text{where}\,\, \omega_{k-1}^{m,p} = \frac{\sqrt{(k - 1)!(k - 1 -m)!}}{(k - 1 - p)!}}
\end{equation}
 \n (ii) By using the main result of  {\color{blue} [16]}, it was  showed in {\color{blue} [17]} that the operators  $\displaystyle{\mathbb{H}}$ and $\displaystyle{\mathbb{H} + \mathbb{H}^{*}}$ are chaotic operators in Devaney sense where $\displaystyle{\mathbb{H}^{*}}$ is adjoint of $\displaystyle{\mathbb{H}}$ given by $\displaystyle{\mathbb{H}^{*}e_{k} = \omega_{k}^{{p,m}}e_{k+1}}$ .\\
 
 \end{remark}
 
\n {\color{red}\Large{{\bf References}}}\\

\n {\color{blue}[1]}   Ando, T. and Zerner, M., Sur une valeur propre d'un op\'erateur, Comm. Math. Phys. 93 (1984).\\

\n {\color{blue}[2]} Askey R., Wilson J., Some basic hypergeometric orthogonal polynomials that generalize Jacobi polynomials, Mem. Am. Math. Soc., 54, no.319, (1985), 1-55\\

\n {\color{blue}[3]}  Bargmann, V., On a Hilbert space of analytic functions and an associated integral transform I, Comm. Pure Appl. Math. 14 (1962) 187-214.\\

\n {\color{blue}[4]} Berezanskii, Yu. M., Expansion in eigenfunctions of selfadjoint operators. Providence, RI: Am. Math.Soc., (1968)\\

\n {\color{blue}[5]}  Bleicher, M.N.,  Some theorems on vector spaces and the axiom of choice (Fund. Math. 54 (1964), 95-107).http://matwbn.icm.edu.pl/ksiazki/fm/fm54/fm5419.pdf\\

\n {\color{blue}[6]} Borzov, V.V. , Orthogonal polynomials and generalized oscillators algebras, arXiv:math/0002226v1 [math.CA] 26 Feb 2000\\

\n {\color{blue}[7]}  Decarreau, A., Emamirad, H.,  Intissar,A., Chaoticit\'e de l'op\'erateur de Gribov dans l'espace de Bargmann, C. R. Acad. Sci. Paris , 331, (2000).\\

\n {\color{blue}[8]} Devaney, R.L., An Introduction to Chaotic Dynamical Systems, 2nd edn. Addison-Wesley, Reading (1989).\\

\n {\color{blue}[9]} Hellinger, E., Zur Stieltjesschen Kettenbruchtheorie, Mathematische Annalen, vol. 86, (1922) 18-29.\\

\n {\color{blue}[10]} Intissar, A., Spectral Analysis of Non-self-adjoint Jacobi-Gribov Operator and Asymptotic Analysis of Its Generalized Eigenvectors, Advances in Mathematics (China), Vol.44, (3), (2015) 335- 353 doi: 10.11845/sxjz.2013117b\\

\n {\color{blue}[11]} Intissar, A., On spectral approximation of unbounded Gribov-Intissar operators in Bargmann space. Adv. Math. (China) 46(1), 13-33 (2017)\\

\n {\color{blue}[12]} Intissar, A.,  A Note on the Completeness of Generalized Eigenfunctions of the Hamiltonian of Reggeon Field Theory in Bargmann Space, Complex Analysis and Operator Theory (2023) https://doi.org/10.1007/s11785-023-01395-z\\

\n {\color{blue}[13]} Intissar, A.,  Etude spectrale d'une famille d'op\'erateurs non-sym\'etriques intervenant dans la th\'eorie des champs de Reggeons, Comm. Math. Phys. 113 (1987) 263-297.\\

\n {\color{blue}[14]}  Intissar, A., Analyse de Scattering d'un op\'erateur cubique de Heun dans l'espace de Bargmann, Comm. Math. Phys. 199 (1998) 243-256.\\
 
\n {\color{blue}[15]} Intissar, A., On a chaotic weighted shift $\displaystyle{z^{p} \frac{d^{p+1}}{dz^{p+1}}}$ of order $p$ in Bargmann space, Advances in Mathematical Physics, (2011), Article ID 471314, 11 pages.\\

\n {\color{blue}[16]} Intissar, A., Intissar, J.K., On chaoticity of the sum of chaotic shifts with their adjoints in Hilbert space and applications to some weighted shifts acting on some Fock-Bargmann spaces. Complex Anal. Oper. Theory volume 11, issue 3, 491-505 (2017)\\

\n {\color{blue}[17]} Intissar, A., Intissar, J.K., A Complete Spectral Analysis of Generalized Gribov-Intissar's Operator in Bargmann Space, Complex Analysis and Operator Theory volume 13, issue 3, pages1481?1510 (2019).\\

\n {\color{blue}[18]} Intissar, A., Le Bellac, M. and Zerner, M., Properties of the Hamiltonian of Reggeon field theory, Phys. Lett. B 113 (1982) 487-489.\\

\n {\color{blue}[19]} Jentzsch, P., Ober Integralgleichungen mit positivem Kern. J. Reine Angew. Math. 141, 235-244 (1912)\\

\n {\color{blue}[20]} Camosso, S. Gaussian integrals depending by a quantum parameter in finite dimensionarXiv:2107.06874v1 [math.GM] 14 Jul (2021)\\

\n {\color{blue}[21]} Kostyuchenko, A.G.,  Mirzoev, K.F., Three-term recurrence relations with matrix coefficients. The completely indeterminate case, Math. Notes, vol. 63, No. 5, (1998) 624-630.\\

\n {\color{blue}[22]} Kostyuchenko, A.G.,  Mirzoev, K.F., Complete indeterminateness tests for Jacobi matrices with matrix entries, Functional Analysis and its applications, vol 35, No. 4, (2001) 265-269.\\

\n {\color{blue}[23]} Kouba, O.,  Inequalities related to the error function. \\https://doi.org/10.48550/arXiv.math/0607694 (2006)\\

\n \n {\color{blue}[24]} Krein, M. G., Infinite $J$-matrices and the matrix moment problem, Dokl.Akad. Nauk SSSR, 69, no. 3 (1949) 125-128\\

\n {\color{blue}[25]} Krein, M.G., The fundamental propositions of the theory of representation of Hermitian operators with deficiency index (m, m), AMS Translations, Ser. 2, vol. 97 (1970) 75-144\\

\n {\color{blue}[26]} Krein, M.G., Rutman, M.A.: Linear operators leaving invariant a cone in a Banach space. Usp. Mat. Nauk (N.S.) 3, 1 (1948) [Am. Math. Soc. Transl. 26, 3-95 (1960)]\\

\n {\color{blue}[27]} Osipov, A.S., On the Hellinger theorem and lp properties of solutions of difference equations. Journal of Difference Equations and Applications, vol 9, No. 9 (2003) 841-851.\\

\n {\color{blue}[28]} Schmdgen , K., Unbounded Self-adjoint Operators on Hilbert Space , Springer, Netherlands, (2012).\\

\end{document}